\documentclass[twoside]{mhd}

\usepackage{setspace}
\usepackage{graphicx}
\usepackage{amsmath}

\usepackage{tikz}
\usepackage{graphicx}
\usetikzlibrary{positioning}

\numberwithin{equation}{section}

\usepackage{graphicx}
\newcommand\Ha{\mbox{\textit{Ha}}}  
\newcommand\Pran{\mbox{\textit{Pr}}} 
\newcommand\Nu{\mbox{\textit{Nu}}} 
\newcommand\Rayl{\mbox{\textit{Ra}}}  

\mhdhead{56}{0}{1}	

\title{Rayleigh-B\'{e}nard Convection in Strong Vertical Magnetic Field: Flow Structure and Verification of Numerical Method}

\author{R. Akhmedagaev\inst{1}, O. Zikanov\inst{1}, D. Krasnov\inst{2}, J. Schumacher\inst{2}}

\institute{University of Michigan - Dearborn, Dearborn, USA
\and Technische Universit\"{a}t Ilmenau, Ilmenau, Germany}

\begin{document}
\maketitle

\begin{abstract}

Direct numerical simulations are performed to study turbulent Rayleigh-B\'{e}nard convection in a vertical cylindrical cavity with uniform axial magnetic field. Flows at high Hartmann and Rayleigh numbers are considered. The calculations reveal that, similarly to the behavior observed in Rayleigh-B\'{e}nard convection  with strong rotation, flows at strong magnetic field develop a central vortex, while the heat transfer is suppressed.

\end{abstract}

\section{Introduction}

Rayleigh-Bénard convection in the presence of a magnetic field plays a significant role in a plethora of natural and industrial processes where it affects the turbulent transport of heat and momentum \cite{Weiss:2014,Ozoe:2005}. The classical picture of the magnetic field effects includes suppression of turbulence and increase of the critical Rayleigh number $\Rayl_c$. In the case of free-slip boundaries, the latter effect is given by \cite{Chandr:1961}
\begin{equation} \label{eq:Ra_cr}
{\Rayl_c} = \frac{\pi^2 + a^2}{a^2} \left[ (\pi^2 + a^2)^2 + \pi^2\Ha^2\right] 
\end{equation} where $a$ = $kH$, $H$ and $\Ha$ are the dimensionless horizontal normal mode wavenumber, the height of the convection layer and the Hartmann number correspondingly. 

The asymptotic behavior $\Rayl_c \approx \pi^2\Ha^2$  also holds for no-slip boundary conditions at the top and bottom of a system. Interestingly, it has been recently found that magnetoconvection is present for $\Rayl_c < \Rayl$ in the form of subcritical modes attached to the sidewalls (the wall modes). Numerical calculations \cite{Liu:2018} of flows in a wide square cell reveal the existence and the complex two-layer structure of these modes conducted at a moderate $\Rayl = 10^7$ and a low Prandtl number $\Pran = 0.025$. These structures are still present for Hartmann numbers up to a doubled value of the linear stability limit $\Ha_c \equiv \sqrt{\Rayl}/{\pi} \approx1000$ and the results show that similarly to experiments on rotating Rayleigh-Bénard system, significant transport of heat and momentum is maintained by subcritical modes which are attached to the sidewalls \cite{Ecke:1993}.

Several experiments, most notably \cite{Cioni:2000}, indicate that the classical picture of the magnetic field suppressing the flow and, thus, reducing the rate of transport is not always correct. In particular, in systems with sidewalls, a very strong magnetic field may lead to growth of the transport rate and the Nusselt numbers higher than in flows with weak or zero magnetic field at the same $\Rayl$.

The ultimate goal of our study is to explore the hypothesis that the enhancement of the heat transfer at high $\Ha$ is caused by the wall modes having the form of ascending/descending jets located near the sidewalls. The results in this paper present the first stage of the study, focused on the flow regimes in critical and moderately supercritical region.


\section{Presentation of the problem}  

\subsection{Physical model}
We consider a flow of an incompressible viscous electrically conducting fluid (liquid metal) with constant physical properties contained in a cylinder with a uniform axial magnetic field. The top and bottom walls are maintained at constant temperatures. The lateral wall is thermally insulated. All walls are perfectly electrically insulated. Using the Boussinesq and quasi-static approximations we write the non-dimensional governing equations as
\begin{equation} \label{eq:1}
\nabla \cdot\bvec{u}  =  0,
\end{equation}
\begin{equation} \label{eq:2}
\frac{\partial \bvec{u}}{\partial t} +
(\bvec{u}\cdot \nabla)\bvec{u} = -\nabla p +  {\sqrt\frac{\Pran}{\Rayl}} (\nabla^2 \bvec{u} +{\Ha}^2\left(\bvec{j} \times \bvec{e}_z\right)) + T\bvec{e}_z,
\end{equation}
\begin{equation} \label{eq:3}
\frac{\partial T}{\partial t} + \bvec{u}\cdot\nabla T
 = {\sqrt{\frac{1}{\Rayl\Pran}}} \nabla^2 T,
\end{equation}
\begin{equation} \label{eq:4}
\bvec{j} = -\nabla \phi + (\bvec{u} \times \bvec{e}_z),
\end{equation}
\begin{equation} \label{eq:5}
\nabla^2\phi = \nabla \cdot (\bvec{u} \times \bvec{e}_z),
\end{equation} where $p$, $\bvec{u}$, $\nabla \phi$ and $T$ are the fields of pressure, velocity, electric potential, and deviation of temperature from a reference value.

The governing equations are made dimensionless by using the cylinder's height $H$, the free-fall velocity $U=\sqrt{g \alpha \Delta T H}$, the external magnetic field strength $B_0$  and the imposed temperature difference $\Delta T=T_{bottom}-T_{top}$ as the scale of length, velocity, magnetic field and temperature correspondingly.

The dimensionless control parameters are the Prandtl number $\Pran=\nu/\kappa$, the Rayleigh number $\Rayl=g \alpha \Delta T H^3/\nu \kappa $, the Hartmann number $\Ha=B_0H(\sigma/\rho\nu)^{1/2}$ and the aspect ratio  $\Gamma=D/H$.

\subsection{Numerical method}

Governing equations (\ref{eq:1})$-$(\ref{eq:5}) are solved numerically using the finite difference scheme described earlier in \cite{Krasnov:FD:2011,Krasnov:2012,ZikanovJFM:2013}. The spatial discretization is implemented in cylindrical coordinates with the boundary conditions at the axis specified according to the principles outlined in \cite{Constantinescu:2002} (see \cite{ZikanovJFM:2013} for a discussion). The scheme is of the second order and nearly fully conservative in regards of mass, momentum, kinetic energy, and electric charge conservation principles \cite{Krasnov:FD:2011,Ni1:2007}. The time discretization is semi-implicit and based on the Adams-Bashforth/Backward-Differentiation method of the second order. At every time step, three elliptic equations $-$ the projection method equation for pressure, the equation for temperature (\ref{eq:3}) and for potential (\ref{eq:5}) are solved using the FFT in the azimuthal direction and the cyclic reduction direct solver in the $r-z$ $-$ plane. The computational grid is clustered toward the wall according to the coordinate transformation in the axial direction $z = \tanh(A_z \zeta)/\tanh(A_z)$ and in the radial direction $r = 0.9 \sin(\eta  \pi/2) + 0.1 \eta$ where $-1 \le \zeta \le 1$, $0 \le \eta \le 1$ are the coordinates, in which the grid is uniform. Here we only mention the novel features that appear in the new version of the code. One is the implementation of hybrid (MPI - OpenMP) parallelization in cylindrical coordinates. Another is the new approach to implicit treatment of the viscous terms. In order to avoid the time-step limitations due to diffusive stability limit in narrow grid cells near the axis the Laplacians for three velocity components are discretized implicitly in the azimuthal direction. The resulting one-dimensional equations are solved using the FFT. The radial and axial parts of the Laplacians are treated explicitly. Further details of the numerical method are described in \cite{Krasnov:FD:2011,ZikanovJFM:2013}. 

\subsection{Verification of the model}

The numerical method for our new model has been verified in comparison with the NEK5000 spectral element method (SEM) package for cylindrical coordinates \cite{Schumacher:2016} (without magnetic field) and with the finite-difference (FDM) DNS solver implemented in the Cartesian coordinates, which has been thoroughly verified and applied to high-$\Ha$ flows with and without convection effects in recent studies such as \cite{Liu:2018,Krasnov:2012,Zhang:2014,Zhang:2015,Lv:2014}. Agreement in terms of time-averaged integral characteristics of the flow, such as temperature, kinetic energy and the Nusselt number $\Nu = {\sqrt{\Rayl\Pran}}\langle u_zT \rangle_{A,t} - \langle \frac{dT}{dz} \rangle_{A,t}$ where $\langle \cdot \rangle_{A,t}$ stands for averaging over time and horizontal cross-section, has been achieved. 

\begin{figure}[htb]
\centering

\hskip0.10\textwidth\textbf{$\Rayl = 10^6$} 
\hskip0.25\textwidth  \textbf{$\Rayl = 10^7$}  \\

\begin{tikzpicture}

\node (img1) {\includegraphics[scale=0.185]{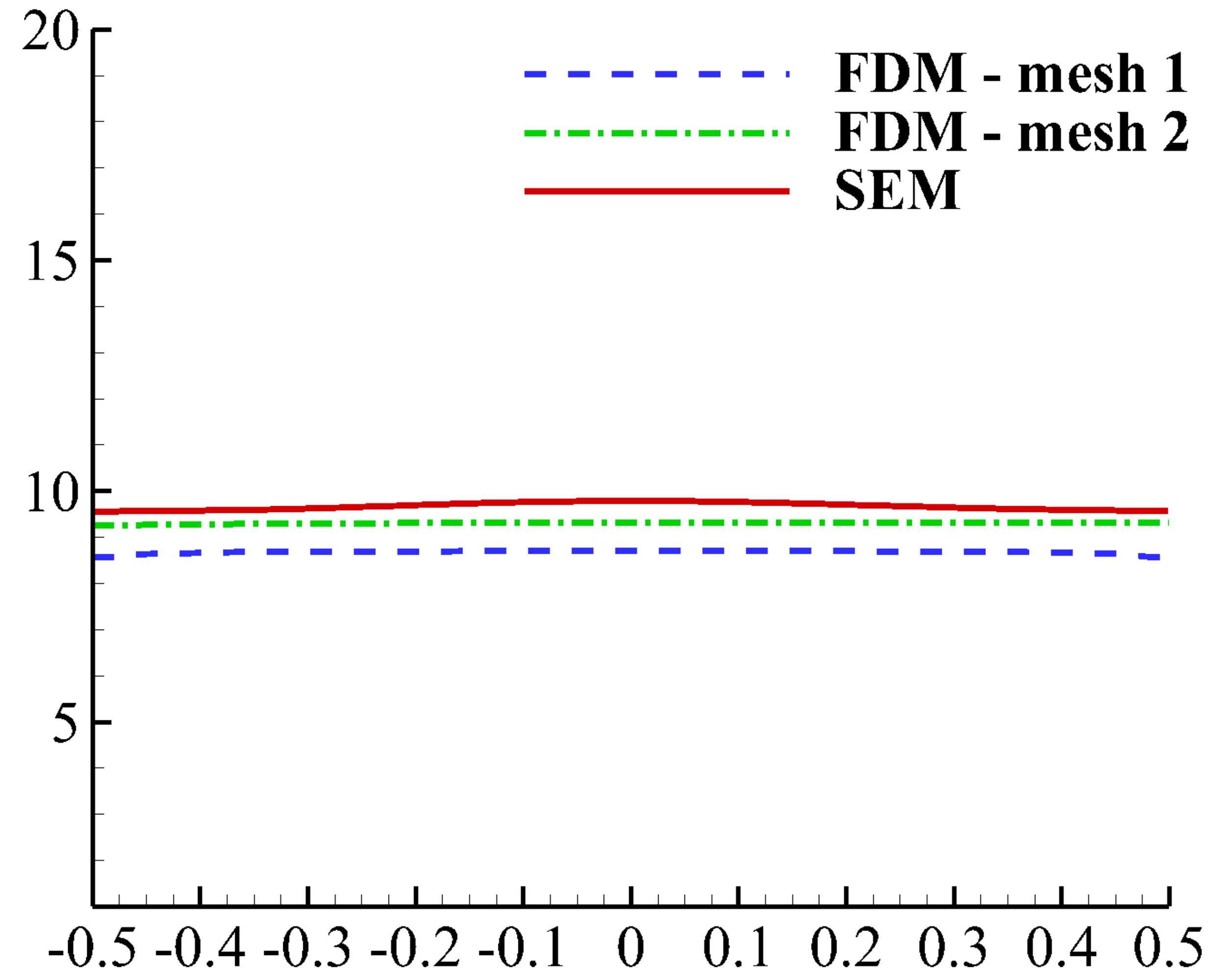}};
	\node[below=of img1, xshift=0.1cm, yshift=1.15cm,font=\color{black}] {\it z}; 
  	\node[left=of img1, xshift=0.85cm ,yshift=1cm,rotate=90,font=\color{black}] {$\langle Nu(z) \rangle_{A,t}$};

\node[right=of img1, xshift=-1cm, yshift=0cm] (img2)  {\includegraphics[scale=0.185]{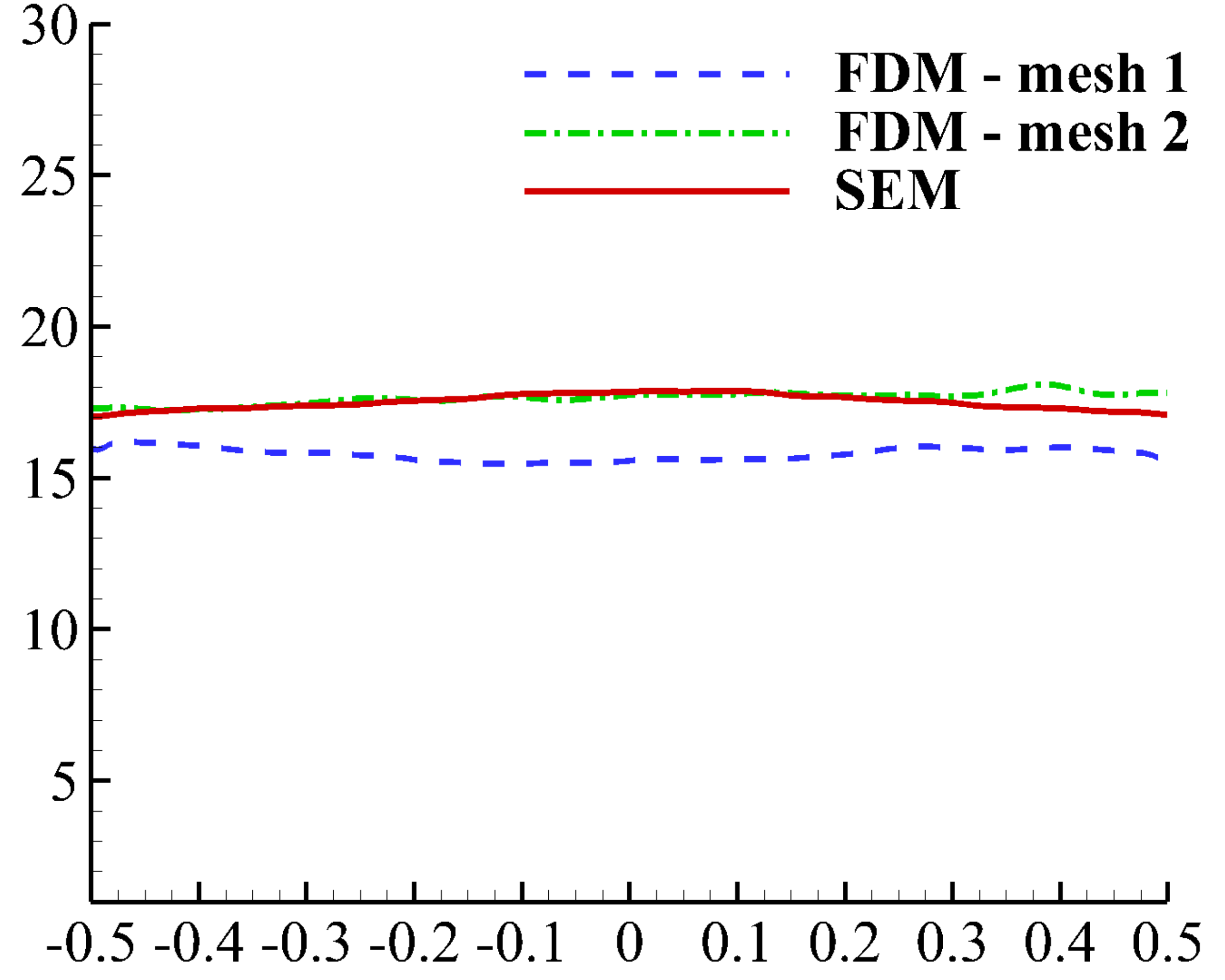}};
	\node[below=of img2, xshift=0.1cm, yshift=1.15cm,font=\color{black}] {\it z}; 
	
	\node[left=of img2, xshift=1.25cm ,yshift=2cm,font=\color{black}] {$ (a) $};
 
\end{tikzpicture}

\begin{tikzpicture}
\hspace*{-0.15cm} 
\node (img1) {\includegraphics[scale=0.183]{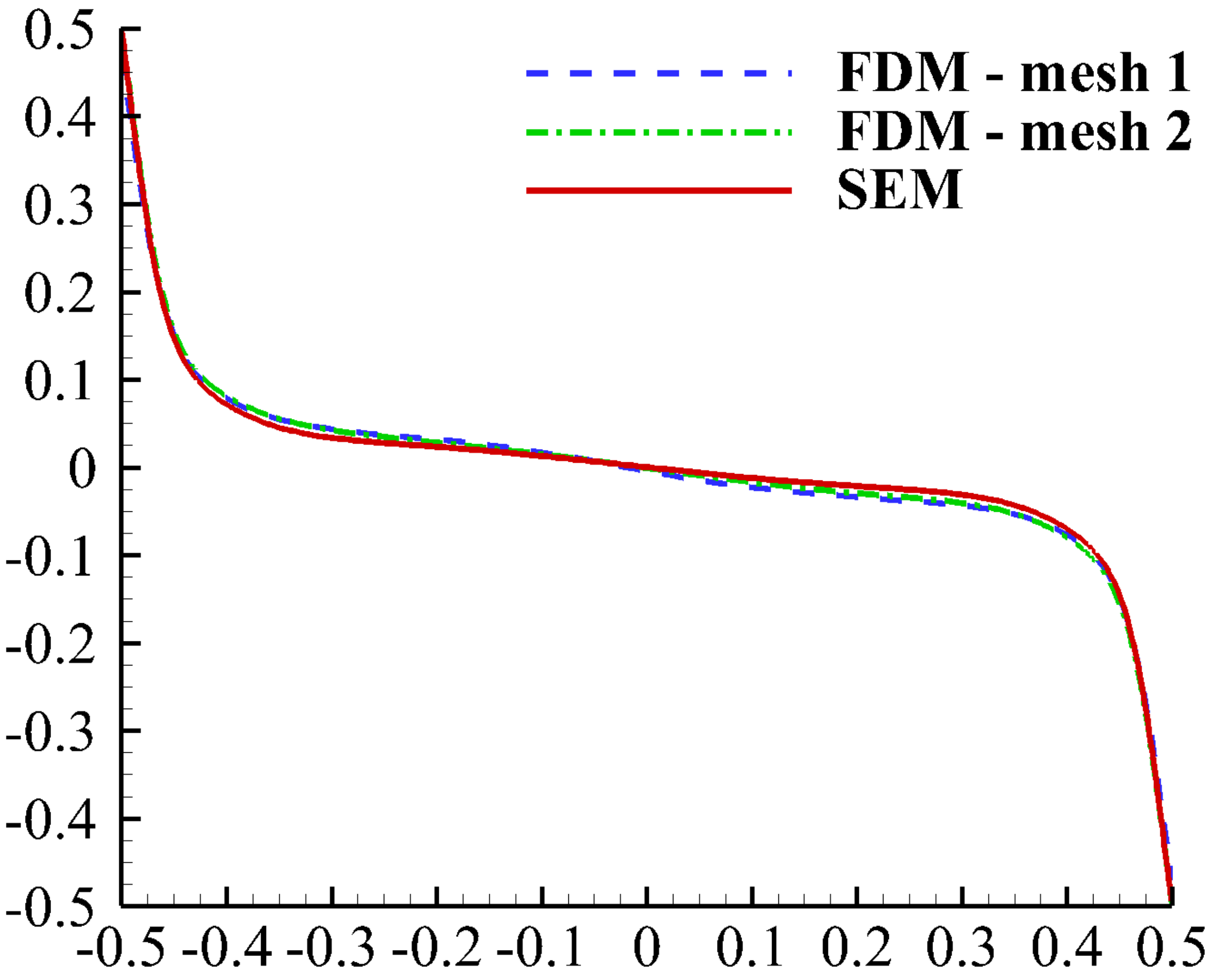}};
	\node[below=of img1, xshift=0.1cm, yshift=1.15cm,font=\color{black}] {\it z}; 
  	\node[left=of img1, xshift=0.85cm ,yshift=1cm,rotate=90,font=\color{black}]{$\langle T(z) \rangle_{A,t}$};

\node[right=of img1, xshift=-1cm, yshift=0cm] (img2)  {\includegraphics[scale=0.183]{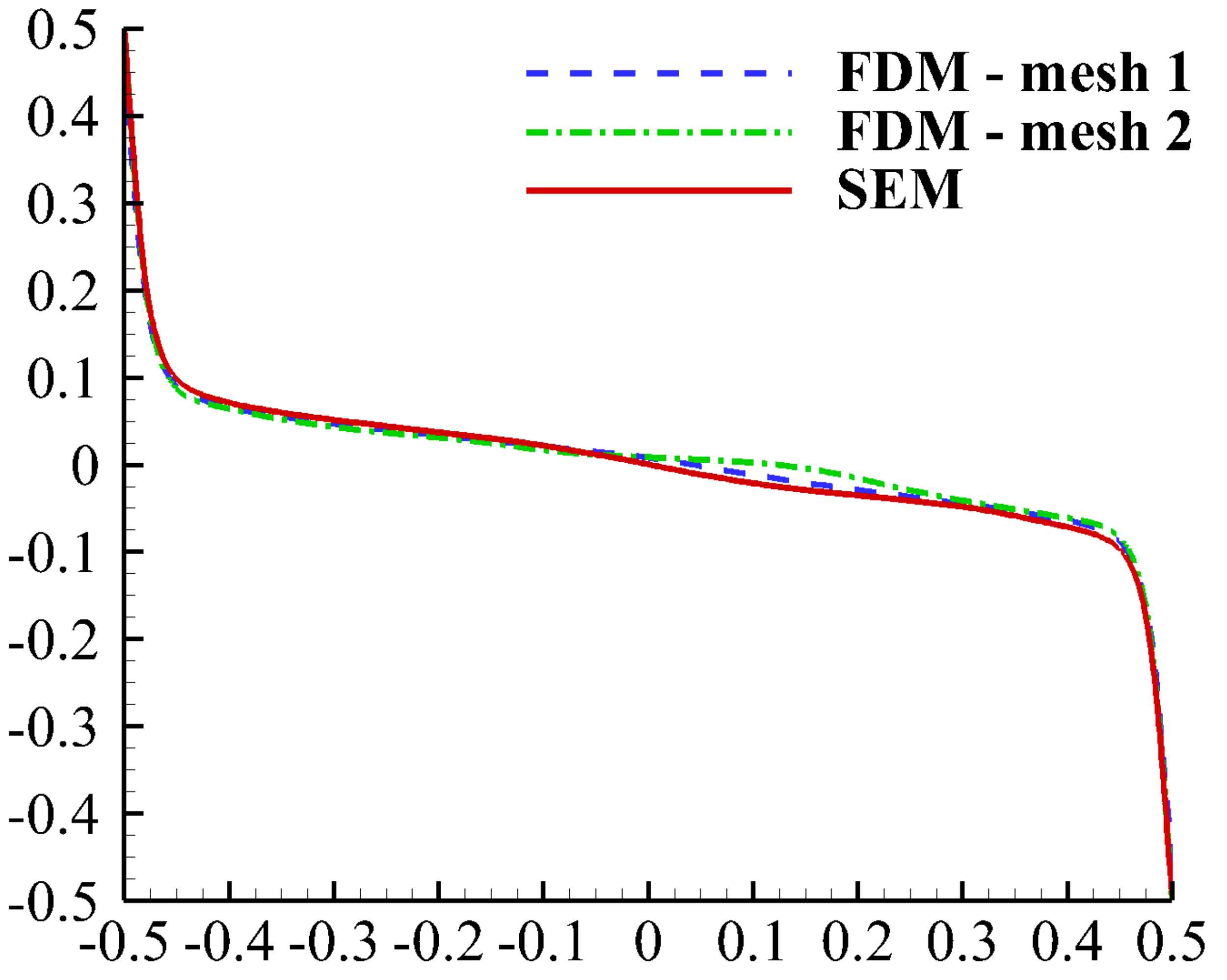}};
	\node[below=of img2, xshift=0.1cm, yshift=1.15cm,font=\color{black}] {\it z}; 
 	
	\node[left=of img2, xshift=1.25cm ,yshift=2cm,font=\color{black}] {$ (b) $};
	
\end{tikzpicture}

\begin{tikzpicture}

\hspace*{0.1cm} 
\node (img1) {\includegraphics[scale=0.185]{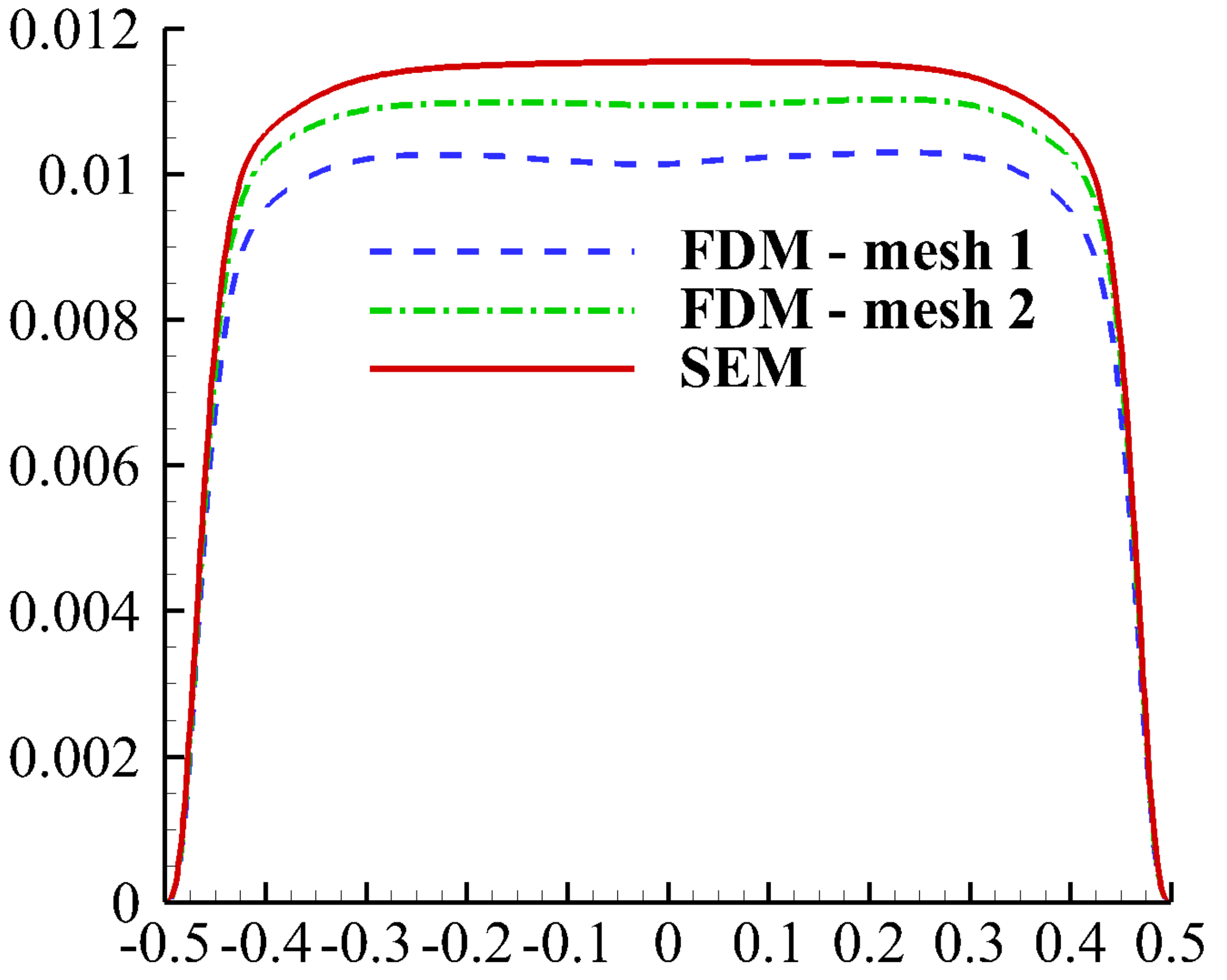}};
	\node[below=of img1, xshift=0.1cm, yshift=1.15cm,font=\color{black}] {\it z}; 
  	\node[left=of img1, xshift=0.85cm ,yshift=1cm,rotate=90,font=\color{black}] {$\langle u_zT(z)\rangle_{A,t}$};
	
\hspace*{-0.75cm}
\node[right=of img1, xshift=-0.25cm, yshift=0cm] (img2)  {\includegraphics[scale=0.185]{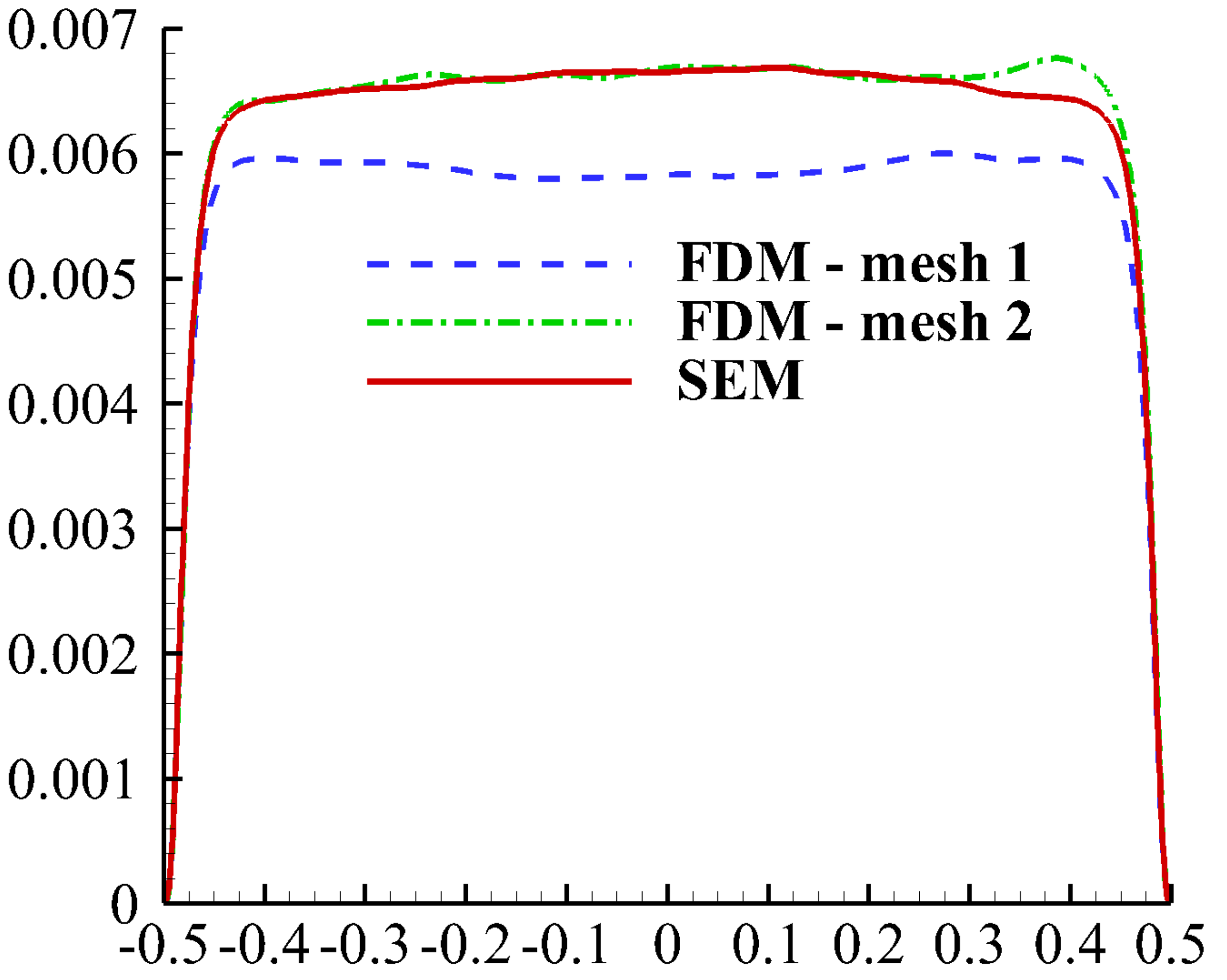}};
	\node[below=of img2, xshift=0.1cm, yshift=1.15cm,font=\color{black}] {\it z}; 
  	
	\node[left=of img2, xshift=1.25cm ,yshift=2cm,font=\color{black}] {$ (c) $};
	
\end{tikzpicture}

\begin{tikzpicture}
\hspace*{-0.125cm} 
\node (img1) {\includegraphics[scale=0.185]{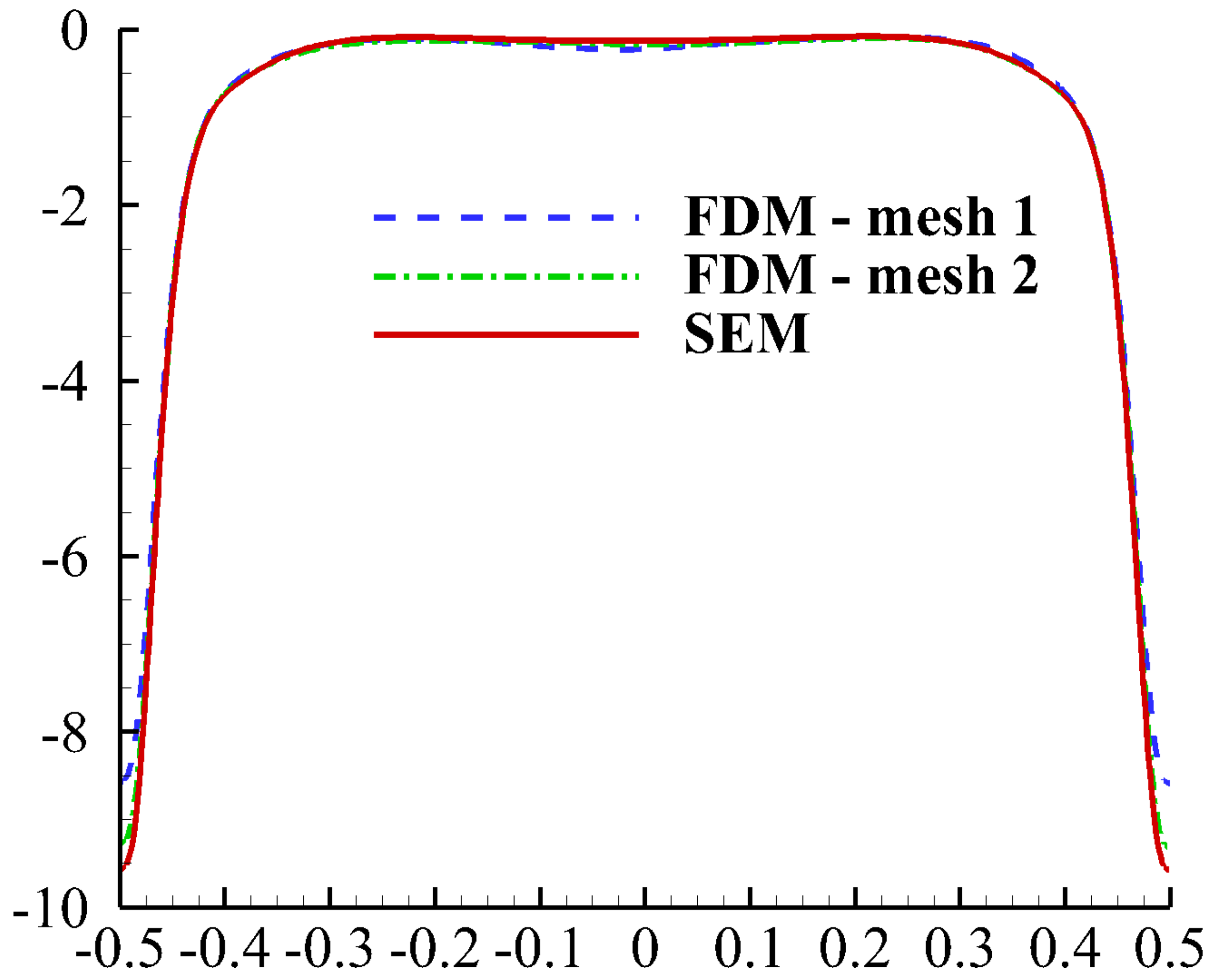}};
	\node[below=of img1, xshift=0.1cm, yshift=1.15cm,font=\color{black}] {\it z}; 
  	\node[left=of img1, xshift=0.85cm ,yshift=1cm,rotate=90,font=\color{black}] {$\langle \frac{dT}{dz}(z) \rangle_{A,t}$};

\node[right=of img1, xshift=-1cm, yshift=0cm] (img2)  {\includegraphics[scale=0.185]{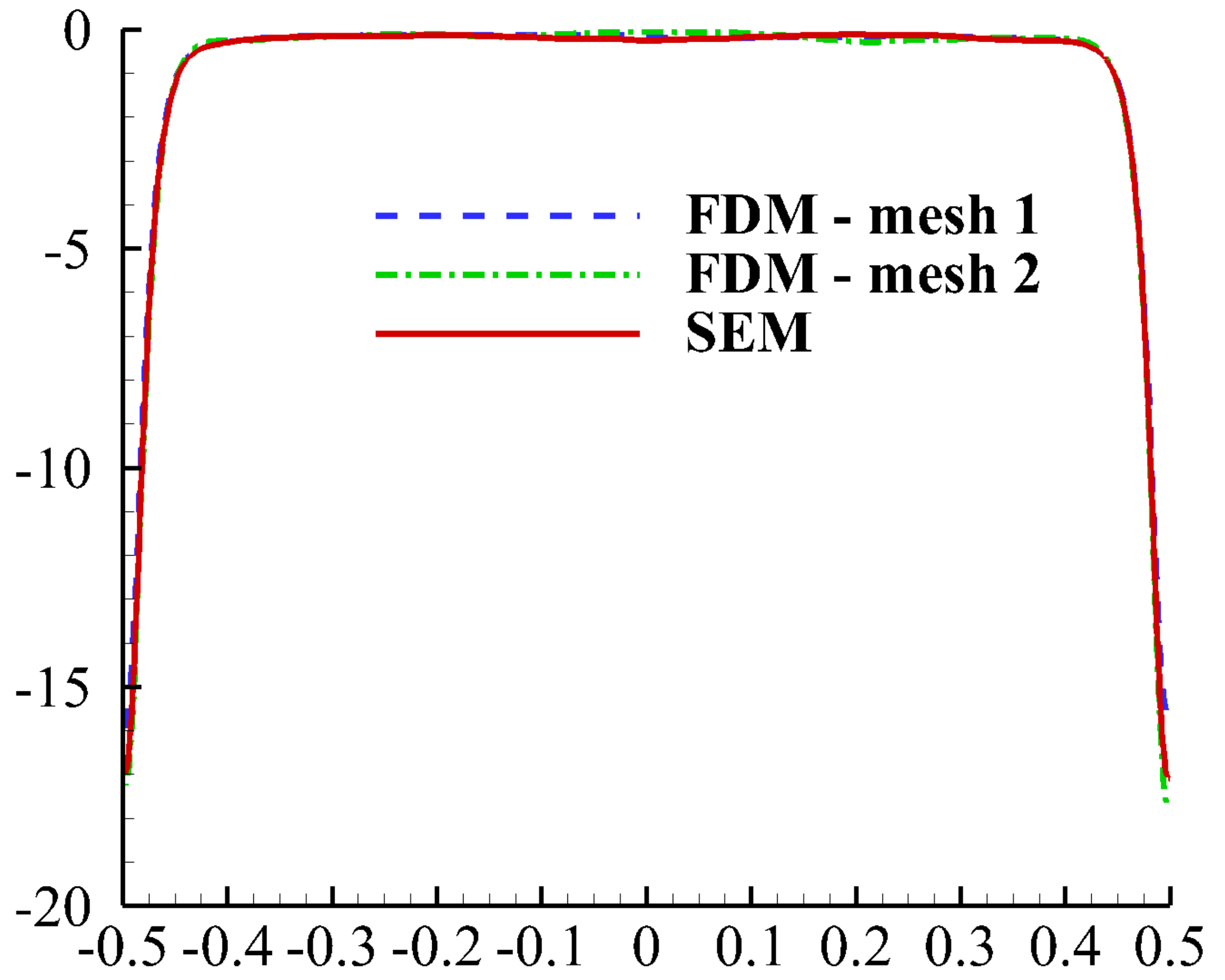}};
	\node[below=of img2, xshift=0.1cm, yshift=1.15cm,font=\color{black}] {\it z}; 
  	
	\node[left=of img2, xshift=1.25cm ,yshift=2cm,font=\color{black}] {$ (d) $};
	
\end{tikzpicture}

\caption{Results of simulations for the vertical mean profiles of the Nusselt number \textbf{\it{(a)}}, temperature \textbf{\it{(b)}}, the convective flux \textbf{\it{(c)}} and the diffusive heat flux \textbf{\it{(d)}} for turbulent flows at $\Pran = 0.7$, $\Ha = 0$, $\Rayl = 10^6$ (left column) and $\Rayl = 10^7$ (right column). SEM \cite{Schumacher:2016} on a grid of $N_e = 61440$ elements with a polynomial order of 5 (left column) and 7 (right column). FDM: mesh 1 - $N_r \times N_z \times N_\theta = 64 \times 128 \times 48$, mesh 2 - $N_r \times N_z \times N_\theta = 192 \times 384 \times 128$ (left column); mesh 1 - $N_r \times N_z \times N_\theta = 64 \times 128 \times 48$, mesh 2 - $N_r \times N_z \times N_\theta = 256 \times 512 \times 192$ (right column).}
\label{fig1}
\end{figure}

\begin{figure}
\centering

\hskip0.10\textwidth\textbf{$\Rayl = 10^6$} 
\hskip0.30\textwidth  \textbf{$\Rayl = 10^7$}  \\

\includegraphics[scale=0.3]{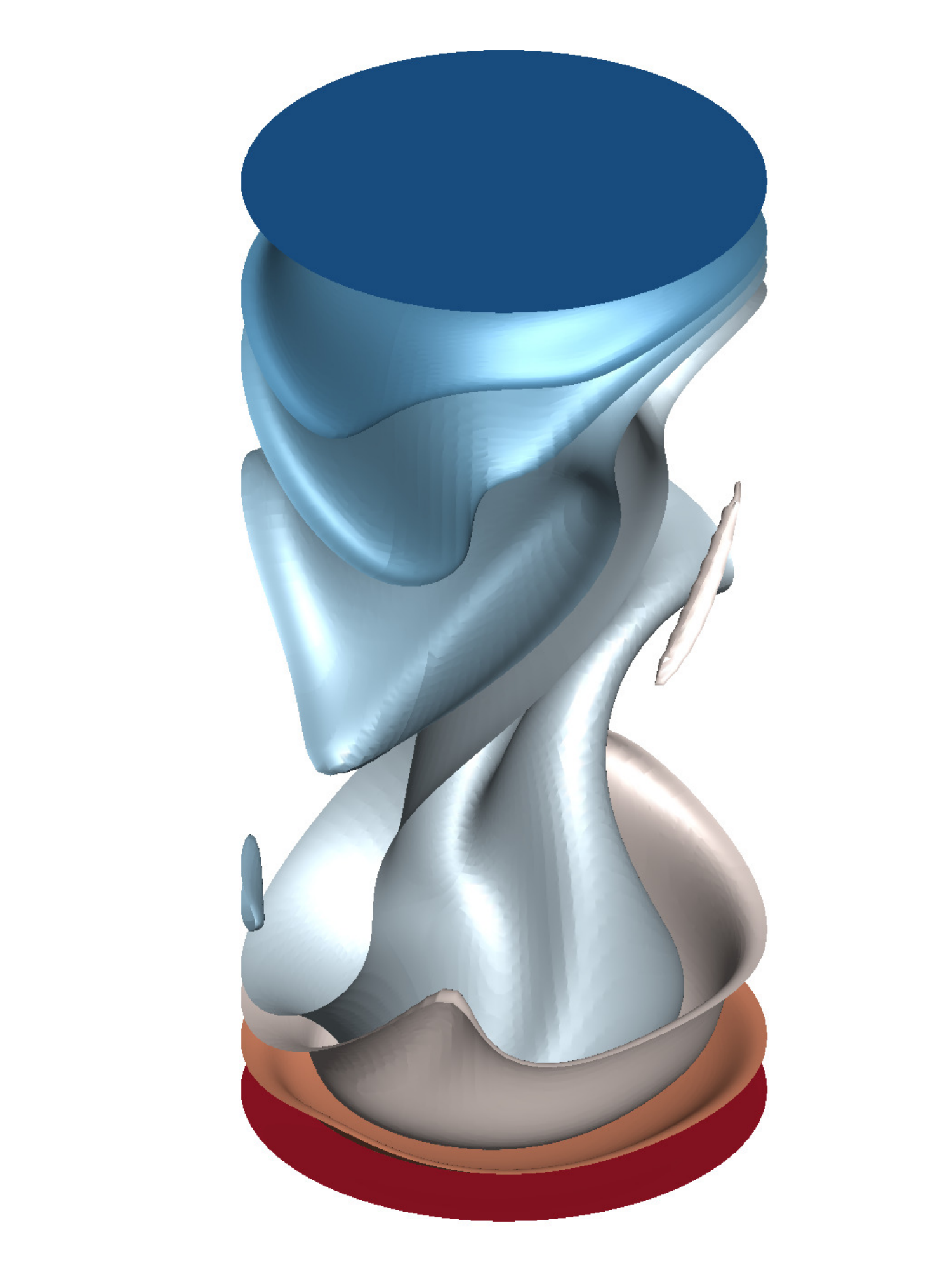}\
\includegraphics[scale=0.3]{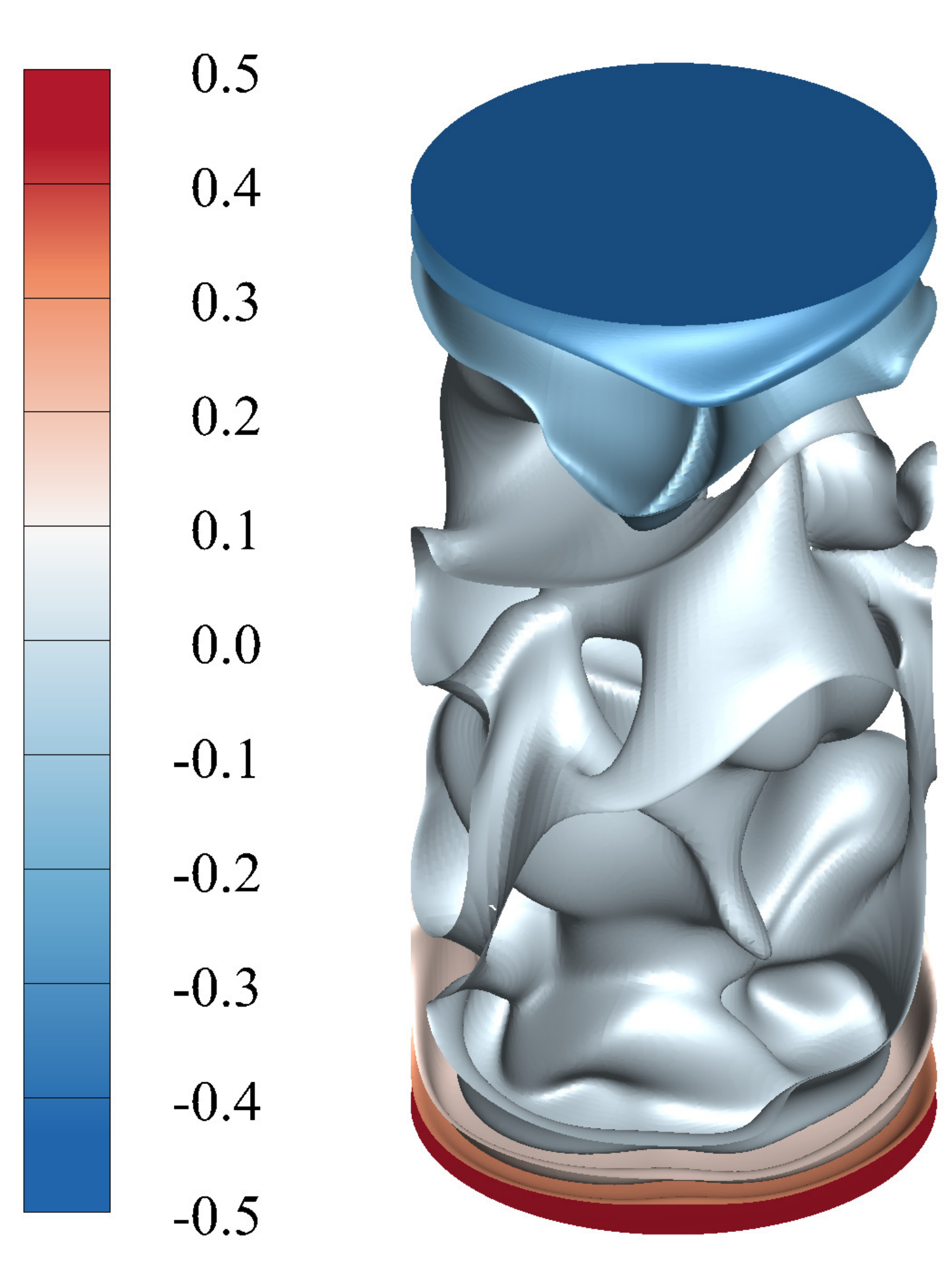}\\

\caption{Instantaneous temperature iso-surfaces for $\Pran = 0.7$ at $\Ha = 0$, $\Rayl = 10^6$ and $\Ha = 0$, $\Rayl = 10^7$. The aspect ratio is $\Gamma=1/2$.}
\label{fig2}
\end{figure}

For comparison with the results obtained by the spectral element method we choose the following control parameters: $\Rayl = 10^6, 10^7$, $\Pran = 0.7$, $\Ha = 0$ and $\Gamma=1/2$. Our simulations at two Rayleigh numbers show that the results obtained with our finite difference scheme converge to those obtained by the spectral element method as the resolution increases. This is true for vertical mean profiles of the Nusselt number $ \langle \Nu(z) \rangle_{A,t}$ (see Fig.~\ref{fig1}a), temperature $ \langle T(z) \rangle_{A,t}$ (see Fig.~\ref{fig1}b), the convective flux $\langle u_zT(z)\rangle_{A,t}$ (see Fig.~\ref{fig1}c) and the diffusive heat flux $\langle \frac{dT}{dz}(z) \rangle_{A,t}$ (see Fig.~\ref{fig1}d). The flow structure can be seen on instantaneous temperature iso-surfaces for two regimes in Fig.~\ref{fig2}.

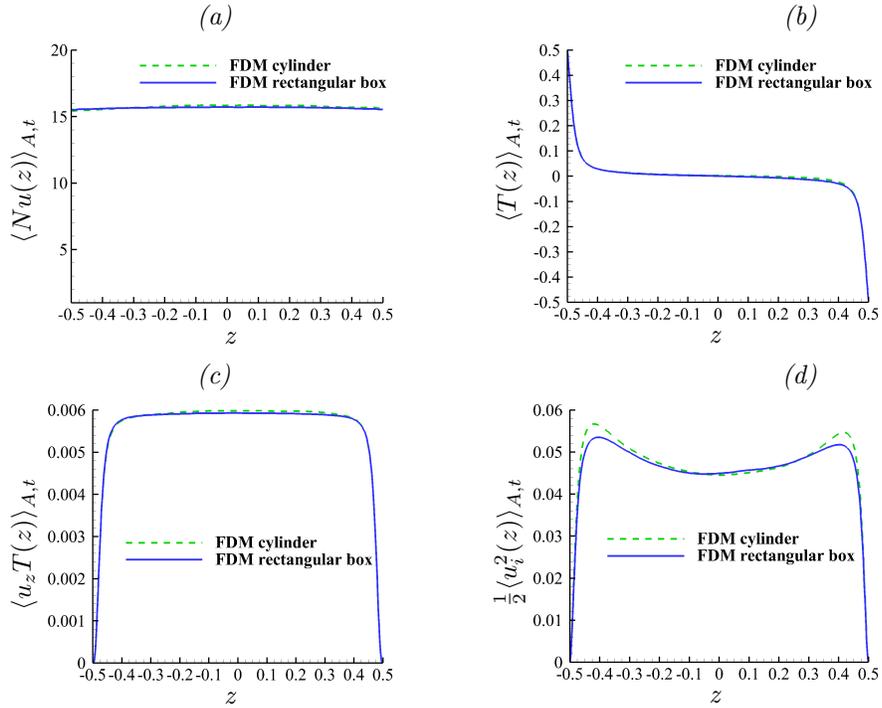
\begin{figure}
\centering

\hskip0.1\textwidth\textbf{\it{(a)}} 
\hskip0.45\textwidth  \textbf{\it{(b)}} \\

\begin{tikzpicture}

\hspace*{0cm}
\node (img1) {\includegraphics[scale=0.185]{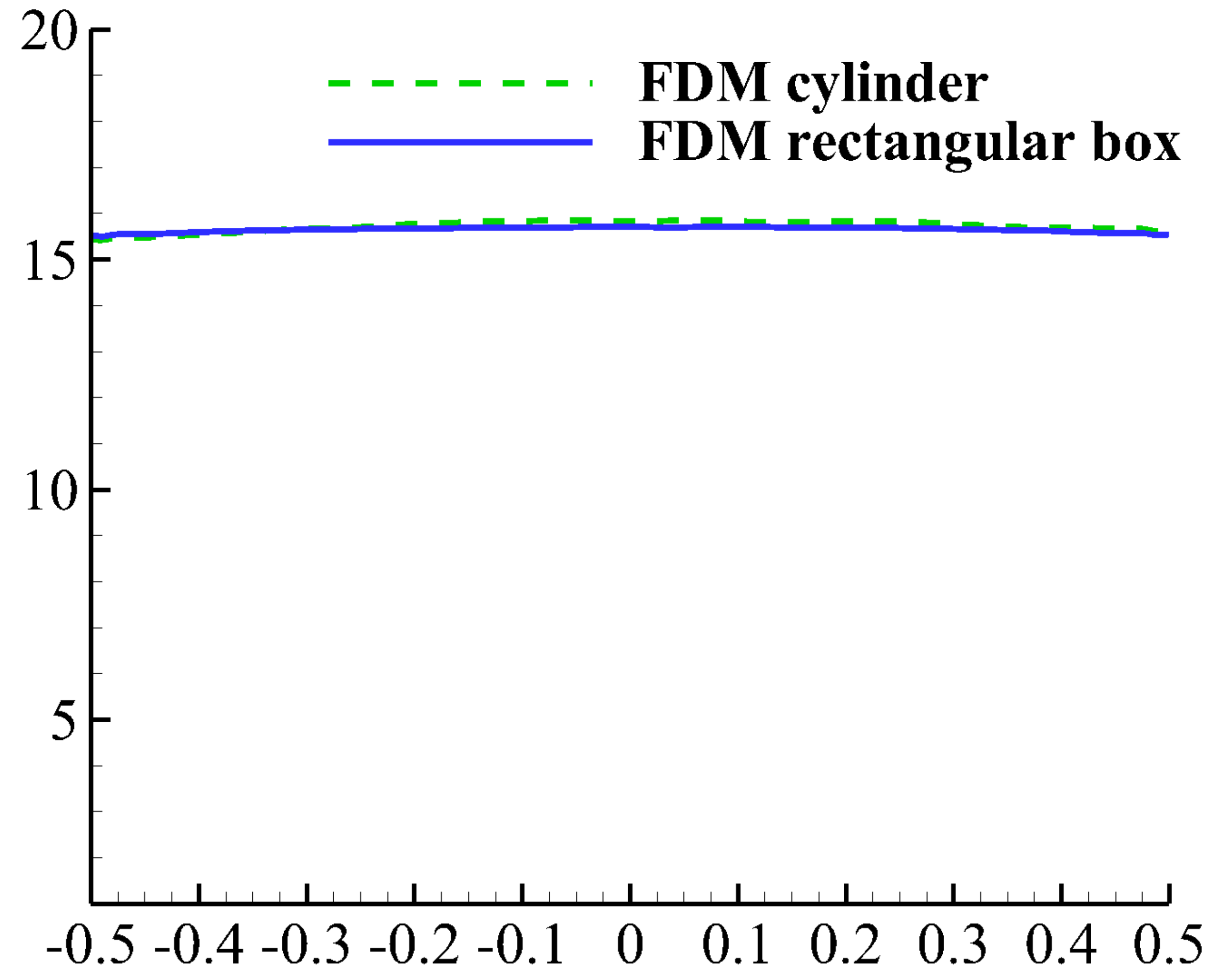}};
	\node[below=of img1, xshift=0.1cm, yshift=1.15cm,font=\color{black}] {\it z}; 
  	\node[left=of img1, xshift=0.85cm ,yshift=1cm,rotate=90,font=\color{black}] {$\langle Nu(z) \rangle_{A,t}$};

\node[right=of img1, xshift=0.5cm, yshift=0cm] (img2)  {\includegraphics[scale=0.185]{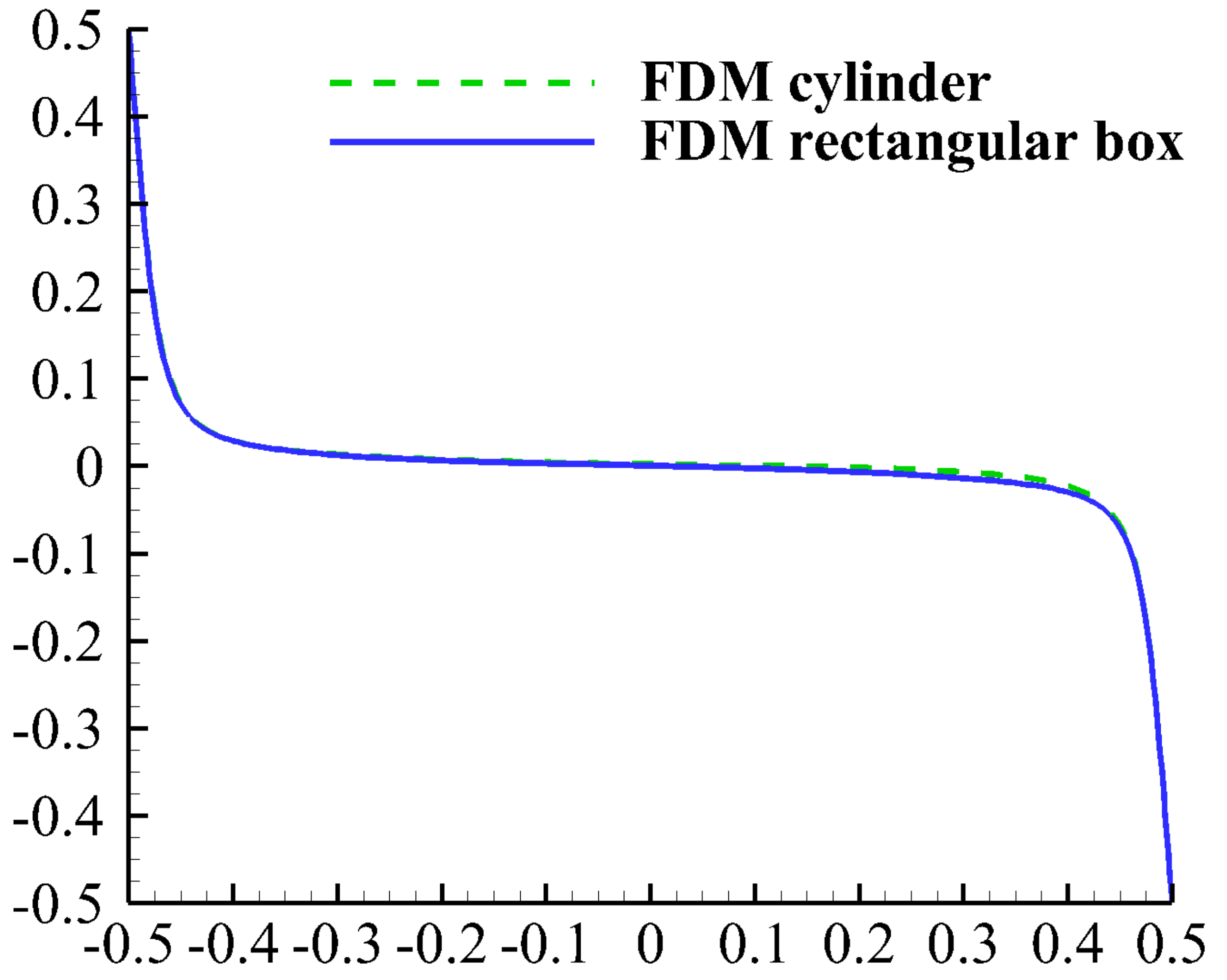}};
	\node[below=of img2, xshift=0.1cm, yshift=1.15cm,font=\color{black}] {\it z}; 
  	\node[left=of img2, xshift=0.85cm ,yshift=1cm,rotate=90,font=\color{black}] {$\langle T(z) \rangle_{A,t}$};
 
\end{tikzpicture}

\hskip0.1\textwidth\textbf{\it{(c)}} 
\hskip0.45\textwidth  \textbf{\it{(d)}} \\

\begin{tikzpicture}

\hspace*{0cm}
\node (img1) {\includegraphics[scale=0.185]{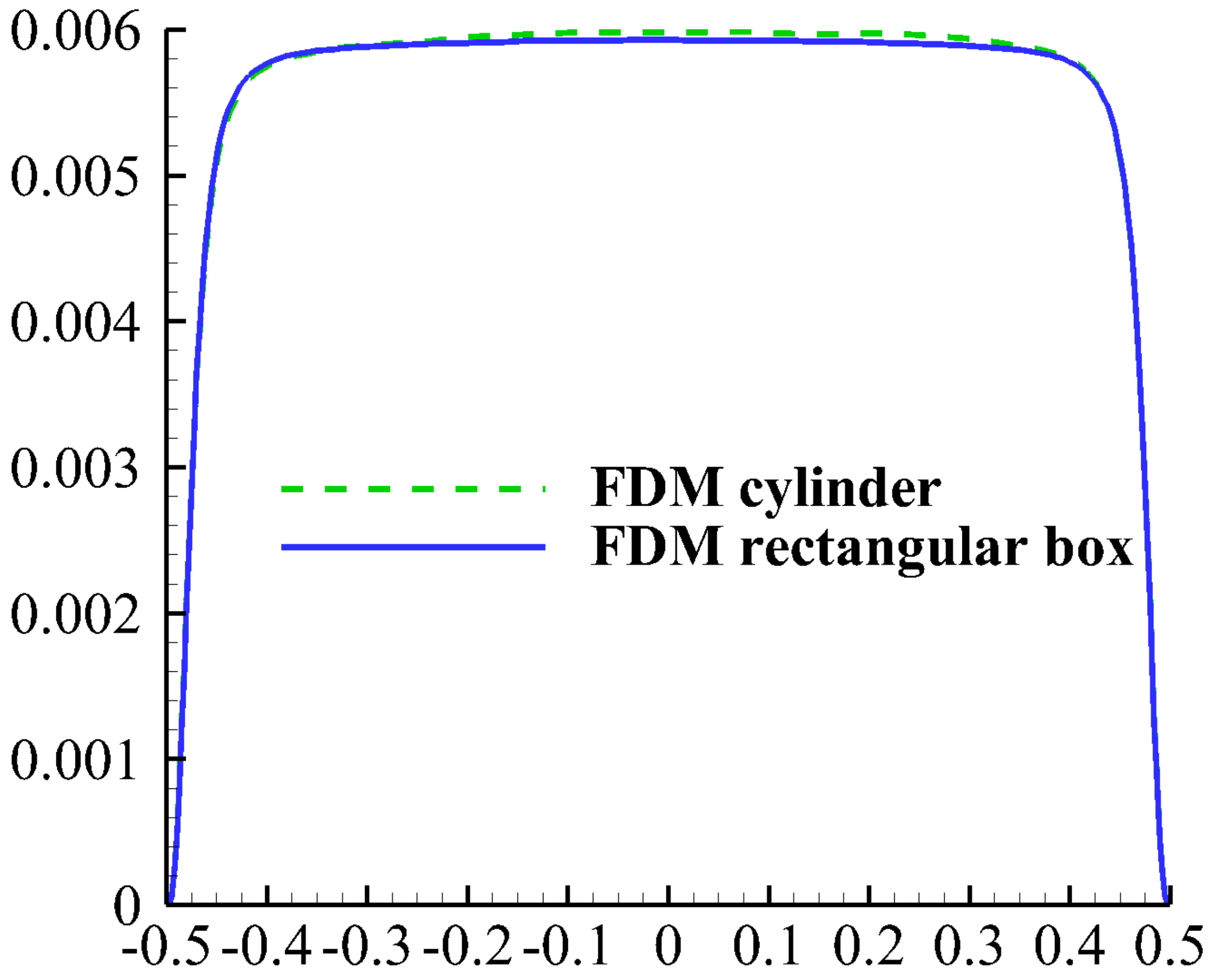}};
	\node[below=of img1, xshift=0.1cm, yshift=1.15cm,font=\color{black}] {\it z};  
  	\node[left=of img1, xshift=0.85cm ,yshift=1cm,rotate=90,font=\color{black}] {$\langle u_zT(z)\rangle_{A,t}$};

\node[right=of img1, xshift=0.5cm, yshift=0cm] (img2)  {\includegraphics[scale=0.185]{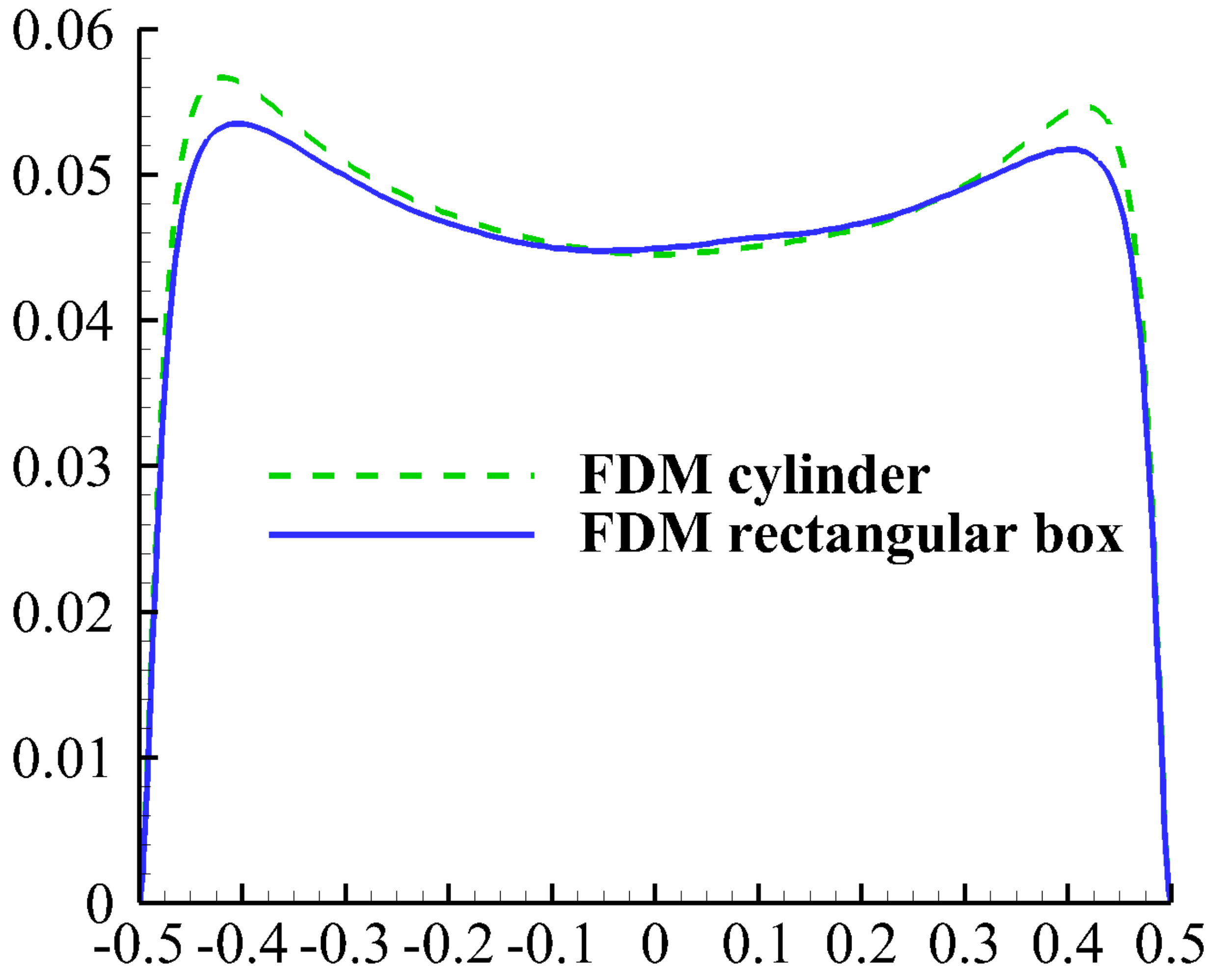}};
	\node[below=of img2, xshift=0.1cm, yshift=1.15cm,font=\color{black}] {\it z}; 
  	\node[left=of img2, xshift=0.85cm ,yshift=1cm,rotate=90,font=\color{black}] {$ \frac{1}{2} \langle u^2_i(z) \rangle_{A,t}$};
 
\end{tikzpicture}

\caption{Results of simulations for the vertical mean profiles of the Nusselt number \textbf{\it{(a)}}, temperature \textbf{\it{(b)}}, the convective flux \textbf{\it{(c)}} and kinetic energy \textbf{\it{(d)}} at $\Ha = 0$, $\Pran = 0.7$, $\Rayl = 10^7$: mesh for a cylindrical cavity - $N_r \times N_z \times N_\theta = 384 \times 192 \times 384$ and for a square cell  - $N_r \times N_z \times N_\theta = 768 \times 192 \times 768$.}
\label{fig3}
\end{figure}

\begin{figure}
\centering

\begin{tikzpicture}
\hspace*{-0.5cm}
\node (img1) {\includegraphics[scale=0.3]{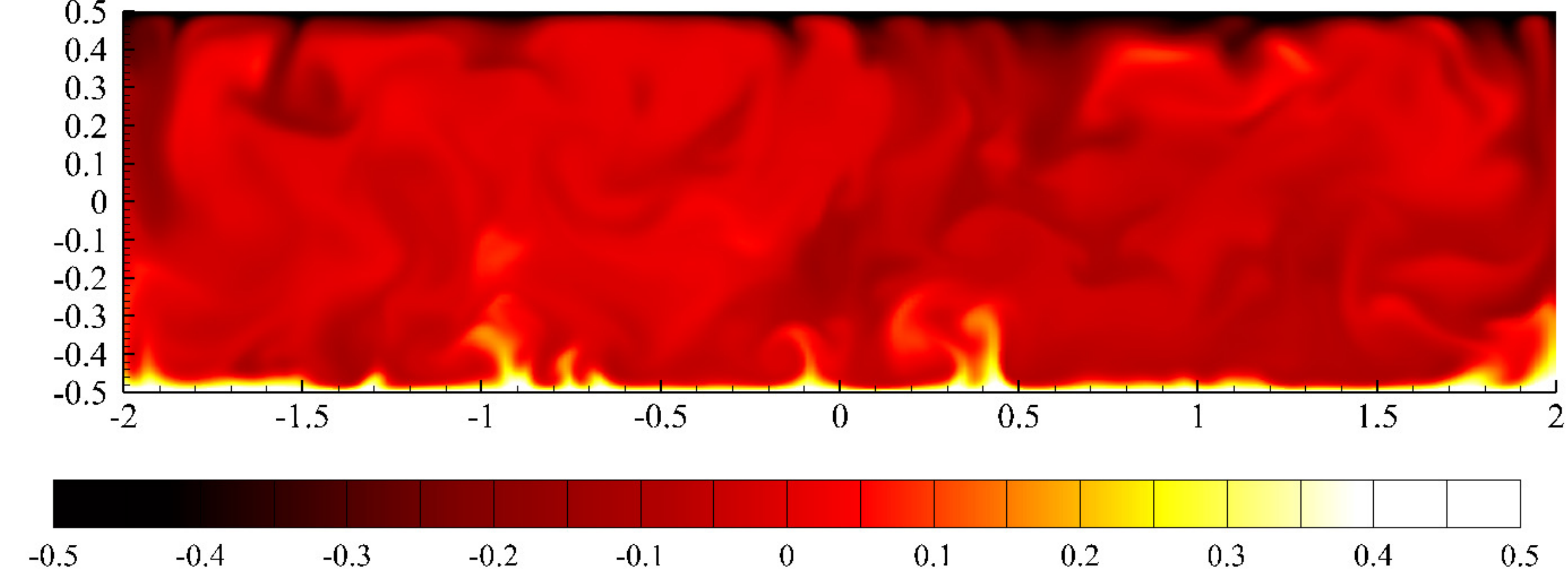}};
	\node[below=of img1, xshift=0.4cm, yshift=2.15cm,font=\color{black}] {\it r}; 
  	\node[left=of img1, xshift=1.25cm ,yshift=0.8cm,rotate=90,font=\color{black}] {\it z};  
\end{tikzpicture}

\caption{Instantaneous temperature distribution in the vertical cross-section through the axis of the cylinder at $\Ha = 0$, $\Pran = 0.7$, $\Rayl = 10^7$. The aspect ratio is $\Gamma=4$.}
\label{fig4}
\end{figure}

As a further verification, comparison was made with the DNS in the Cartesian coordinates performed for a square domain with $\Gamma=4$ \cite{Liu:2018}. We choose the following control parameters $\Rayl = 10^7$, $\Pran = 0.7$ and $\Ha = 0$. We expect that the effect of different geometries becomes insignificant at such large aspect ratio. The results prove that point and present a quite good similarity between the profiles of the Nusselt number (see Fig.~\ref{fig3}a), temperature (see Fig.~\ref{fig3}b), the convective flux (see Fig.~\ref{fig3}c) and kinetic energy $ \frac{1}{2} \langle u^2_i(z) \rangle_{A,t}$ (see Fig.~\ref{fig3}d) obtained in the cylindrical and rectangular geometries. The instantaneous temperature distribution in the vertical cross-section through the axis of the cylinder can be seen in Fig.~\ref{fig4}.

\section{Results}

\begin{figure}
\centering

\hskip0.025\textwidth   \textbf{$\Ha = 0$, $\Rayl/\Rayl_c = \infty$} 
\hskip0.050\textwidth  \textbf{$\Ha = 300$, $\Rayl/\Rayl_c = 12.25$}  \\

\setlength{\parskip}{0.5em}

\includegraphics[scale=0.25]{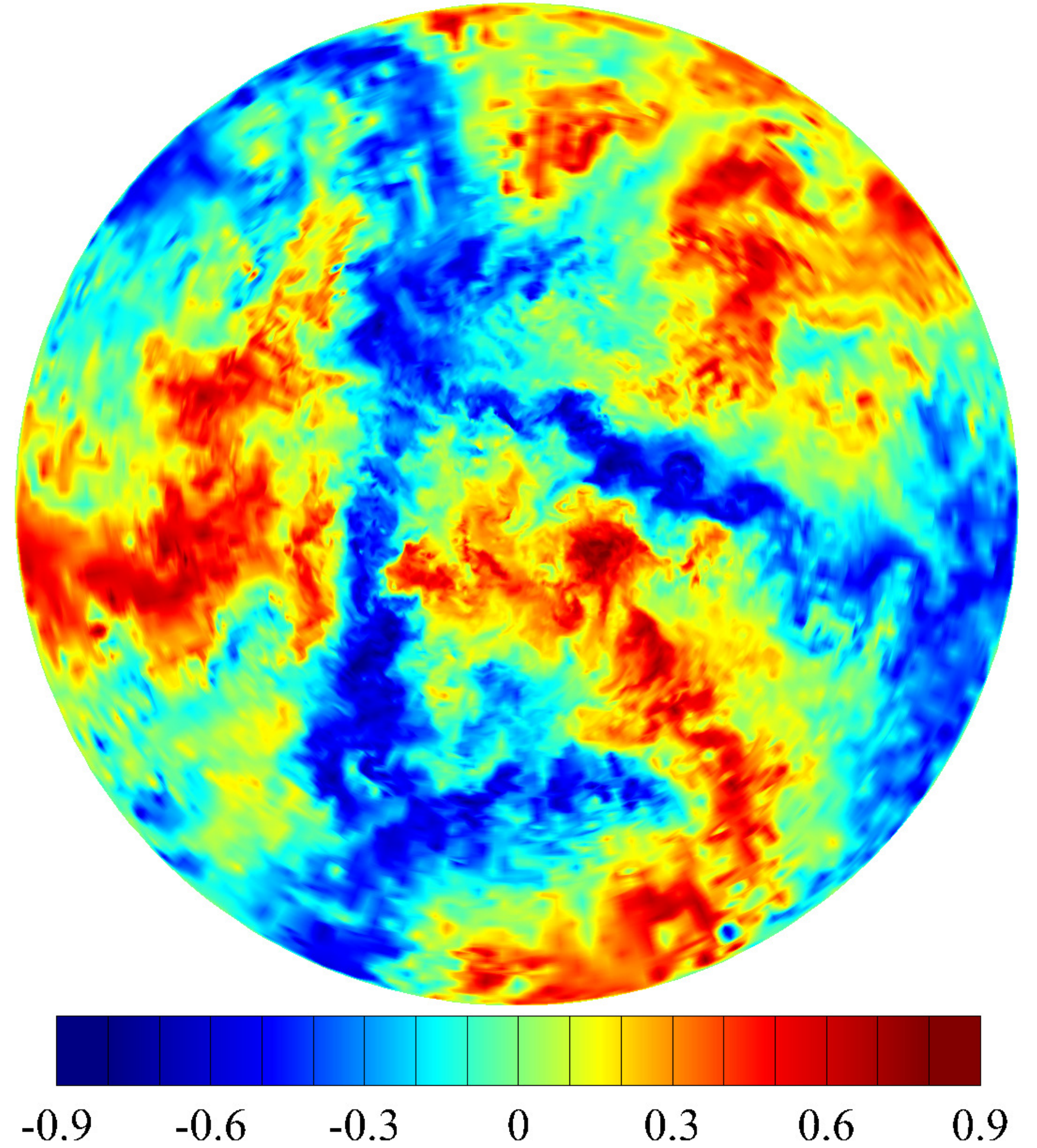}\quad
\includegraphics[scale=0.25]{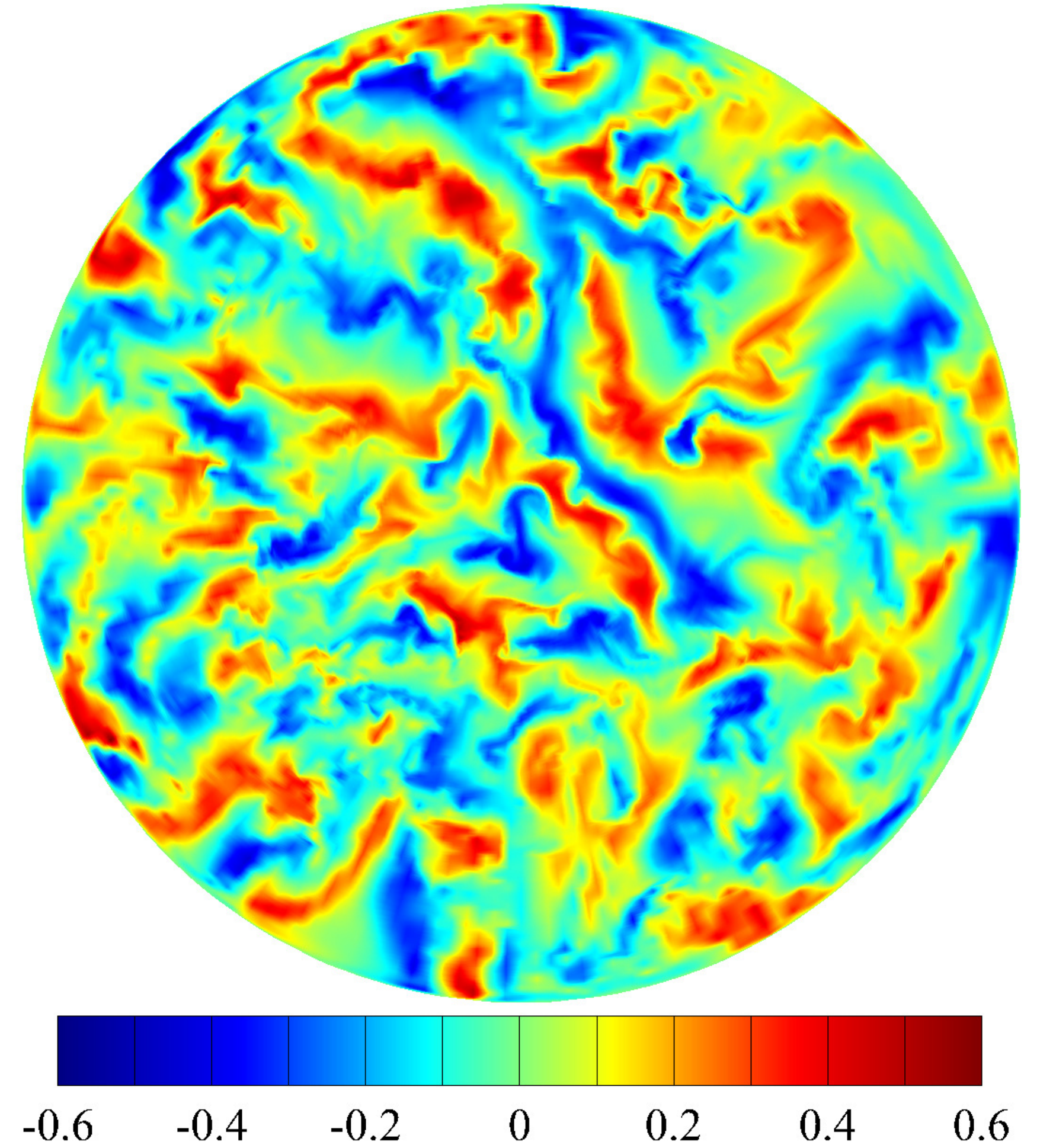}\\

\hskip0.025\textwidth    \textbf{$\Ha = 500$, $\Rayl/\Rayl_c = 4.05$}
\hskip0.050\textwidth    \textbf{$\Ha = 1000$, $\Rayl/\Rayl_c = 1.01$}  \\

\setlength{\parskip}{1em}

\includegraphics[scale=0.25]{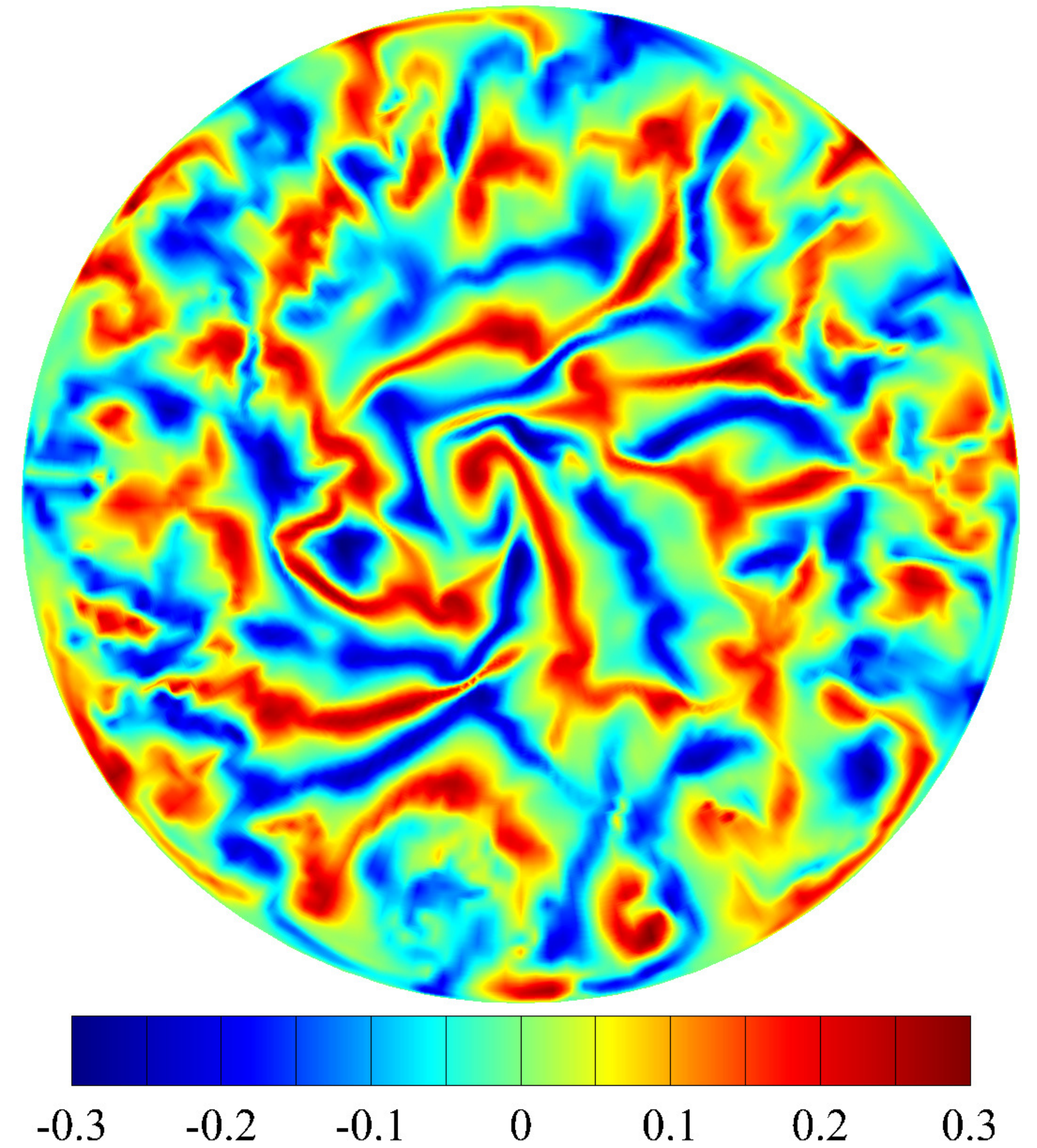}\quad
\includegraphics[scale=0.25]{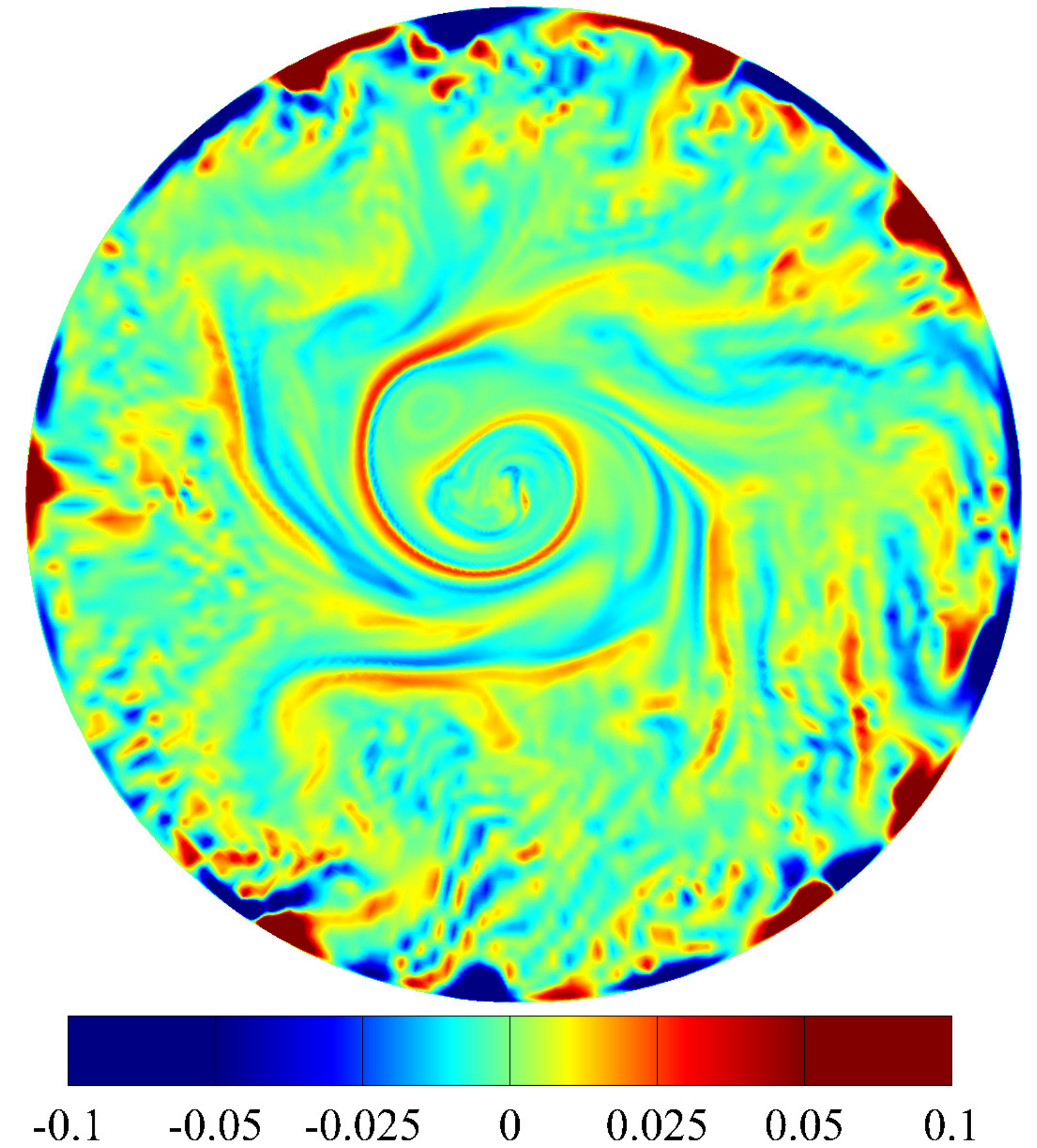}\\

\caption{Instantaneous distributions of the vertical velocity at the mid-plane of the cylinder at $\Ha = 0 - 1000$, $\Pran = 0.025$, $\Rayl = 10^7$. The aspect ratio is $\Gamma=4$.}
\label{fig5}
\end{figure}

\begin{figure}
\centering

\hskip0.025\textwidth   \textbf{$\Ha = 0$, $\Rayl/\Rayl_c = \infty$} 
\hskip0.050\textwidth  \textbf{$\Ha = 300$, $\Rayl/\Rayl_c = 12.25$}  \\

\setlength{\parskip}{0.5em}

\includegraphics[scale=0.25]{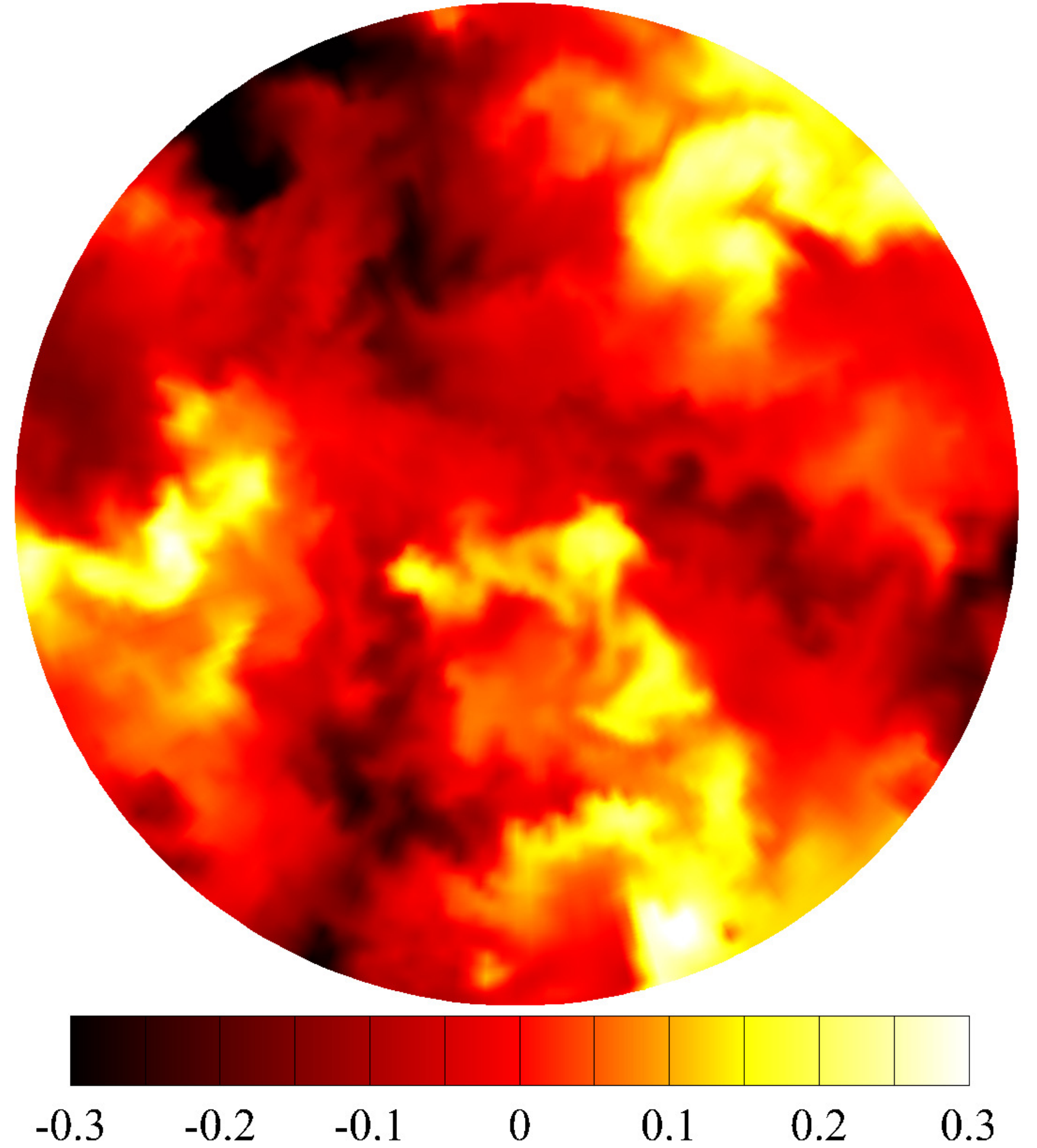}\quad
\includegraphics[scale=0.25]{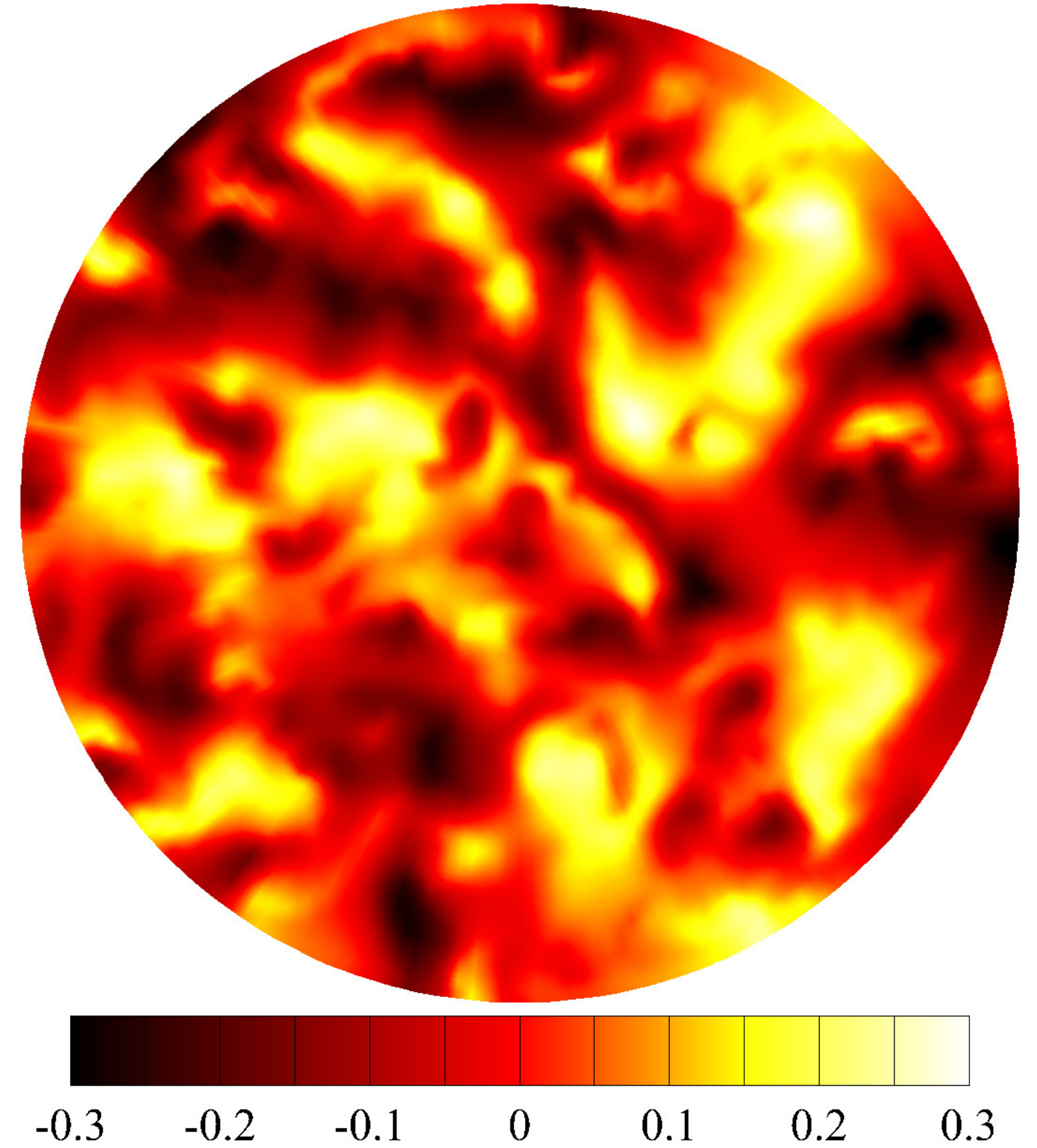}\\

\hskip0.025\textwidth    \textbf{$\Ha = 500$, $\Rayl/\Rayl_c = 4.05$}
\hskip0.050\textwidth    \textbf{$\Ha = 1000$, $\Rayl/\Rayl_c = 1.01$}  \\

\setlength{\parskip}{1em}

\includegraphics[scale=0.25]{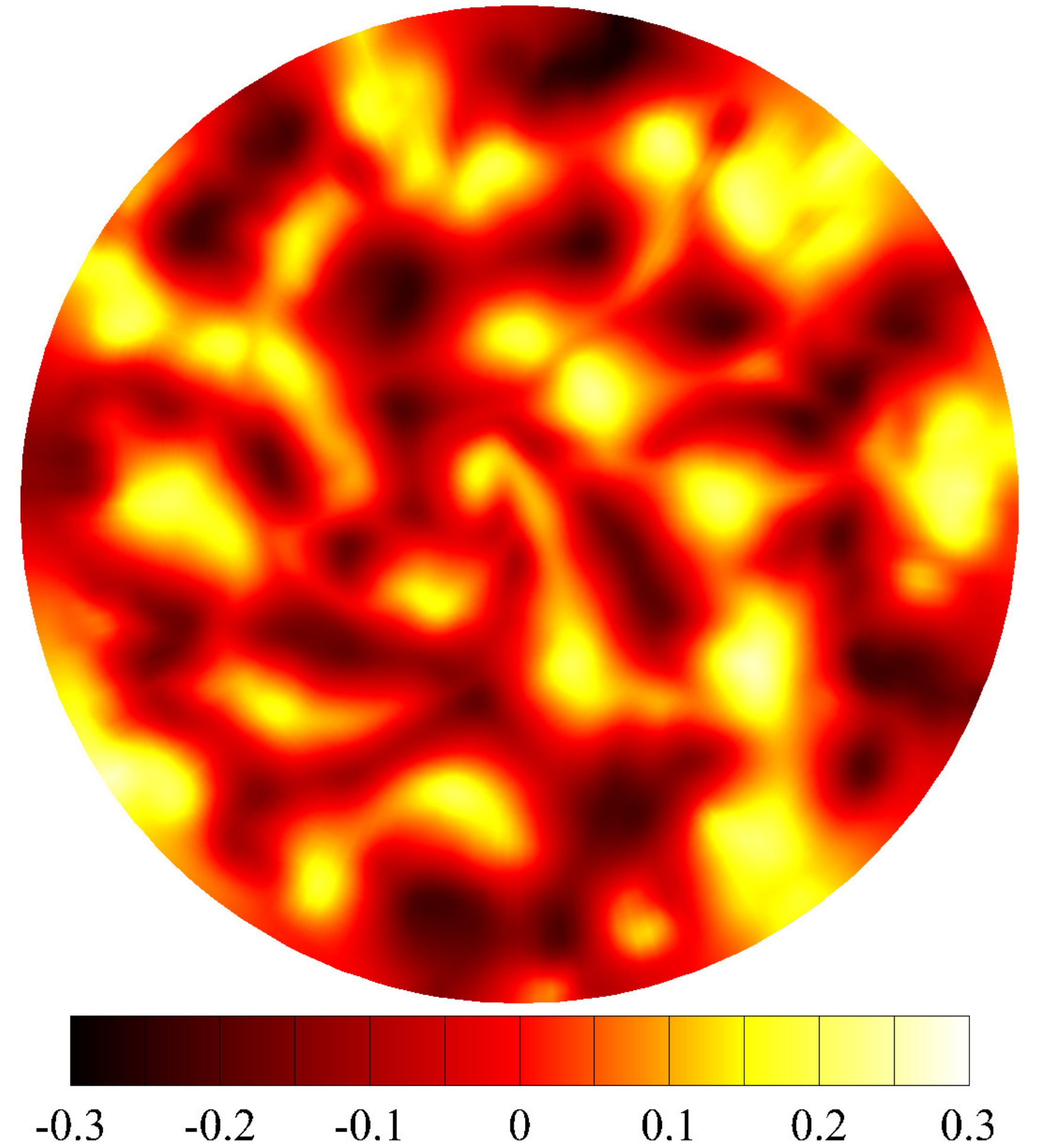}\quad
\includegraphics[scale=0.25]{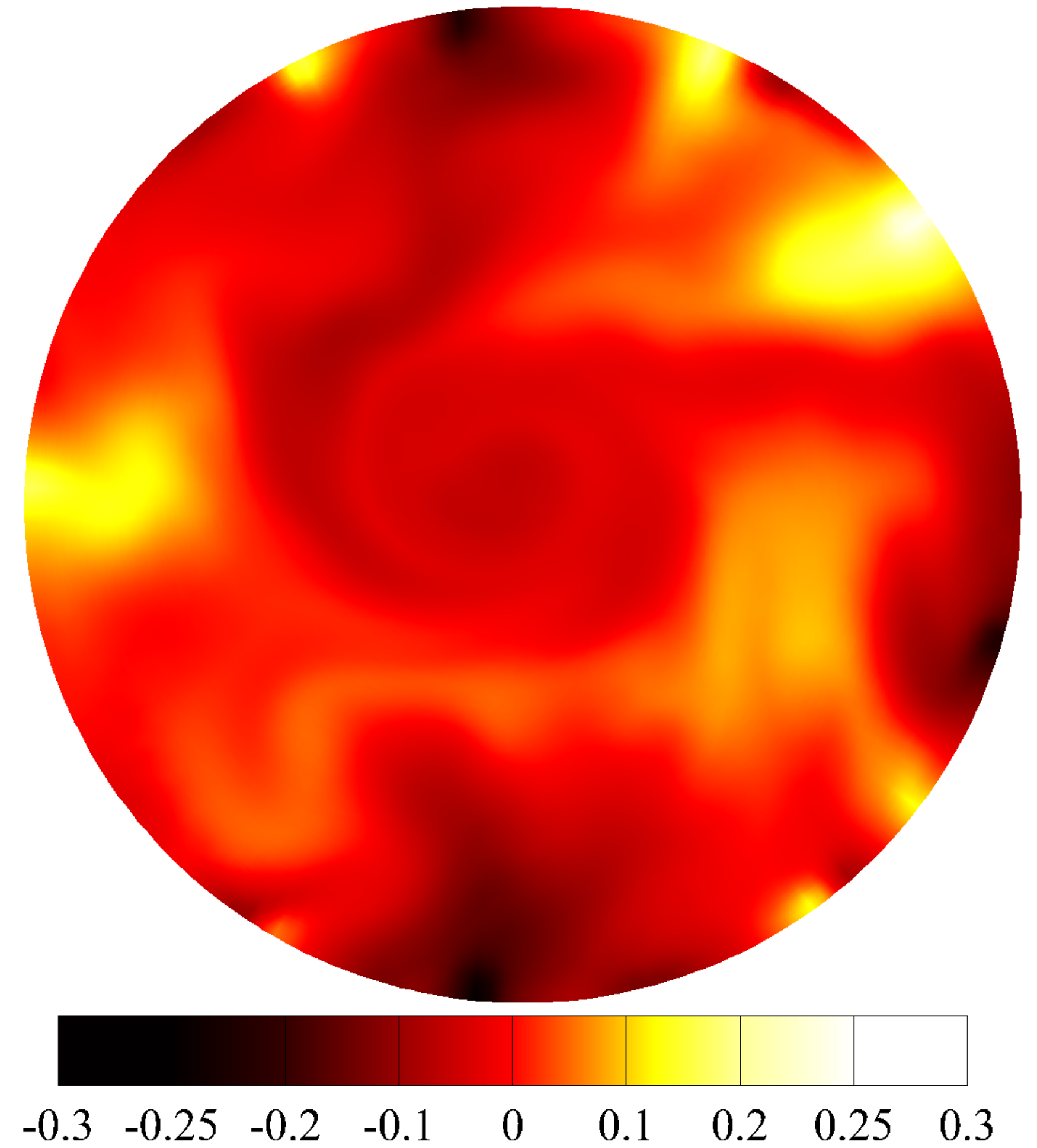}\\

\caption{Instantaneous distributions of temperature at the mid-plane of the cylinder at $\Ha = 0 - 1000$, $\Pran = 0.025$, $\Rayl = 10^7$. The aspect ratio is $\Gamma=4$.}
\label{fig6}
\end{figure}

The focus of the first stage of our work as reported in this paper is on the flow behavior in domains of large aspect ratios at moderately high $\Rayl$. We are exploring the possibility that similarly to rectangular domains \cite{Liu:2018}, flows in cylindrical domains also show subcritical regimes with wall modes near the lateral wall causing enhancement of heat transfer. The computations are conducted for $\Rayl = 10^7$, $\Pran = 0.025$, $\Ha = 0 - 1000$ and $\Gamma=4$. The effects of Rayleigh-B\'{e}nard convection in a cylindrical geometry with large aspect ratios have been studied in \cite{Shishkina:2005,Schumacher:2014,Sakievich:2016}. Current results are for slightly under-resolved DNS with the mesh $N_r \times N_z \times N_\theta = 384 \times 384 \times 384$. They, nevertheless, reveal quite interesting insights which can be seen from the instantaneous distributions of the vertical velocity and temperature at the mid-plane of the cylinder shown in Figs.~\ref{fig5} and \ref{fig6} respectively. 

The flow is turbulent at $\Ha = 0$. An imposed axial magnetic field with $\Ha = 300$ and $\Ha = 500$ creates a cellular structure of up- and downwelling jets that fills the entire bulk region. A similar flow behavior was found in \cite{Liu:2018} for a square cell. Interestingly,  the horizontal velocity shows a vortex near the axis. This is clearly seen at  $\Ha = 500$ in the patterns of $u_z$ and $T$ by the horizontal velocity. In the almost subcritical regime with the ratio of $\Rayl/\Rayl_c = 1.01$ at stronger magnetic field ($\Ha = 1000$), the vortical structure survives in the bulk region in the presence of strong wall modes. The vortex could not been calculated in a square cell due to the difference in geometry. The presence of a central vortex remains an open question for subcritical regimes because the effect of much stronger magnetic fields may totally suppress vortices in the bulk region, preserving only wall modes near the lateral wall.

\section{Conclusions} 

We have developed a new computational model for MHD convection flows in cylindrical geometries based on a highly conservative scheme \cite{Krasnov:FD:2011,Ni1:2007}. The model underwent successful verification based on comparison with two different methods: finite difference and spectral element. 

The advantage of cylinder geometry is that the system possesses an inherent symmetry, due to the periodic direction, which is not the case for rectangular enclosures, considered in the prior studies. This allows for the wall-modes to exhibit an additional degree of freedom, which opens a possibility to address a number of questions relevant to the spatio-temporal evolution of the wall-modes, such as: (i) how will the characteristic wave-number (wave-length) in the circular direction change with the Hartmann number, (ii) will the wall-modes exhibit travelling behavior or remain steady at the walls, (iii) will or will not this behavior be similar to a cylinder convection cell with rotation.

The analysis of the Rayleigh-B\'{e}nard convection in a cylinder at $\Ha = 0 - 1000$, $\Pran = 0.025$, $\Rayl = 10^7$ revealed an expected suppression of small scale fluctuations by increasing magnetic field and, similarly to the prior results of MHD convection in rectangular geometry, residual motion in the form of the wall modes surviving at high Hartmann numbers. The unexpected central vortex develops at high $\Ha$ and its nature remains an open question for future work. 

\Thanks{Financial support is provided by the US NSF (Grant CBET $1803730$) and the DFG grant KR $4445/2 - 1$. Computer time is provided by the Computing Center of the Technische Universit\"{a}t Ilmenau and Super MUC at the LRZ Center.}


\bibliographystyle{mhd}
\bibliography{mhd}



\lastpageno	

\end{document}